\documentclass[lettersize,journal]{IEEEtran}
\usepackage{amsmath,amsfonts}
\usepackage{algorithmic}
\usepackage{algorithm}
\usepackage{array}
\usepackage[caption=false,font=normalsize,labelfont=sf,textfont=sf]{subfig}
\usepackage{textcomp}
\usepackage{stfloats}
\usepackage{url}
\usepackage{verbatim}
\usepackage{graphicx}
\usepackage{booktabs}
\usepackage{tabularx}
\usepackage{arydshln}
\usepackage{hyperref}
\usepackage{multirow}
\usepackage{cite}
\usepackage{xcolor} 

\newif\ifshowrevisioncolor
\ifdefined\cameraready
  \showrevisioncolorfalse
\else
  \showrevisioncolortrue
\fi

\newcolumntype{Y}{>{\centering\arraybackslash}X}
\renewcommand{\arraystretch}{1.1}

\begin{document}

\title{SLM-SS: Token-Based Speech Separation with \\Speech Language Models}

\author{
Chenda Li$^{*,\dagger}$, \textit{Member, IEEE,} Tianhua Li$^{*}$, Yiming Chen, Wei Wang, \\Jiahe Wang, Xin Zhou, Yanmin Qian$^{\dagger}$, \textit{Senior Member, IEEE}\\

\thanks{This work was supported in part by China STI 2030-Major Projects under Grant No. 2021ZD0201500, in part by China NSFC project under Grants No. 62631016, No. 62601524 and No. U25A20409, and in part by SJTU Med-X (Medicine \& Engineering) Translational Research Grant (YG2025LC09). Chenda Li, Tianhua Li, Yiming Chen, Wei Wang, Jiahe Wang, Xin Zhou, and Yanmin Qian are with the Auditory Cognition and Computational Acoustics Lab, Shanghai Jiao Tong University, Shanghai, 200240, P. R. China (\{lichenda1996, herobrine, cym\_3000, wangwei.sjtu, wangjh1030, zxzx818,
yanminqian\}\@sjtu.edu.cn).}
\thanks{$^\dagger$ represents corresponding author, $^*$ represents equal contribution.}}



\maketitle

\begin{abstract}
    Speech separation (SS) has witnessed substantial progress in recent years with the advent of neural network-based methods, achieving notable improvements on signal-level evaluation metrics.
    Existing speech separation systems are mostly optimized with signal-level objectives, and often yield outputs that are acceptable in waveform metrics but suboptimal for intelligibility-sensitive downstream tasks. To address this gap, we propose SLM-SS, a token-based generative speech separation framework that models multi-talker mixtures with speech language models. We first discretize clean speech into multi-codebook token sequences with neural codecs and serialize multi-speaker targets via Serialized Output Training (SOT). An encoder-decoder autoregressive speech language model predicts the zero-order sequence to establish speaker ordering and coarse linguistic content, and a non-autoregressive module then refines higher-order residual codebooks for efficient reconstruction. We further introduce a per-talker non-autoregressive strategy that simplifies high-order modeling and improves the stability.
    On Libri2Mix, SLM-SS with per-talker NAR achieves $6.92\%$ WER, approaching codec-only reconstruction. The WavTokenizer variant eliminates NAR refinement and achieves the highest mean subjective quality score among evaluated separation systems. Its UTMOS and DNSMOS advantages over discriminative baselines persist on unseen noisy and Chinese mixtures without fine-tuning.

\end{abstract}


\section{Introduction}
\label{sec:intro}
\IEEEPARstart{S}{peech} separation (SS) is a fundamental problem in multi-talker speech processing, aiming to isolate individual speech sources from overlapping-signal mixtures~\cite{ss-survey}. It plays a critical role in a wide range of real-world applications, including multi-talker automatic speech recognition (ASR), speaker identification (SID), and assistive technologies such as hearing aids. Currently, SS has been predominantly addressed using discriminative approaches~\cite{ss-survey,deepclustering,deepcasa,conv-tasnet,bsrnn-luo2023,sepformer-cem2021} , which are commonly optimized with signal-level objectives such as scale-invariant signal-to-distortion ratio (SI-SDR)~\cite{sisdr}. Although these methods have demonstrated strong performance in terms of waveform reconstruction and objective evaluation metrics, the separated signals often contain distortions or artifacts, which may have some impact on speech intelligibility and linguistic structure, substantially degrading the performance of downstream tasks such as ASR and SID~\cite{distortion}. In contrast, generative methods~\cite{sepdiff-chenbo,DiffSep-robin2023,scheiblerSourceSeparationFlow2025a,wangMeanSEEfficientGenerative2026}, explicitly model the underlying data distribution of clean speech signals, thereby offering greater potential to produce separated outputs that are more coherent and intelligible. Despite these advantages, existing generative approaches face notable challenges in practice. In particular, the diffusion- or flow-based separation methods~\cite{DiffSep-robin2023,scheiblerSourceSeparationFlow2025a} typically rely on iterative sampling procedures, leading to slow decoding speeds that limit scalability and real-time applicability. Moreover, such models are prone to hallucination~\cite{hallucination}, generating non-existent or spurious speech content, further limiting their reliability in downstream applications.

Meanwhile, recent advances in speech language models (SLMs)~\cite{slmsurvey} have demonstrated strong potential to bridge acoustic and linguistic representations through discrete speech tokens. The SLMs utilize neural codecs~\cite{wavtokenizer,encodec,rvqgan,lisa} to convert continuous speech signals into quantized token sequences. The neural codecs employ residual vector quantization (RVQ)~\cite{grayVectorQuantization1984,vasukiReviewVectorQuantization2006} to compress continuous-space audio into discrete codebook sequences at extremely low bit rates.
Thus, these SLMs enable speech to be processed in an autoregressive next-token prediction manner, allowing powerful sequence modeling techniques originally developed for text language models to be directly applied to speech. As a result, SLM-based approaches have achieved notable progress across a range of intelligibility-oriented speech processing tasks, including ASR~\cite{shiQwen3ASRTechnicalReport2026}, text-to-speech (TTS)~\cite{valle,zhouUniVoiceUnifiedFramework2026}, speech enhancement (SE)~\cite{lanzendorferHighFidelitySpeechEnhancement2026}, conditional target source extraction~\cite{tangTSELMTargetSpeaker2026}, and unified multi-task modeling. In these settings, discrete token representations provide a natural interface for integrating acoustic structure with higher-level semantic information, leading to improved modeling consistency and robustness. Collectively, these developments suggest that discrete SLM-based modeling offers a promising alternative to conventional signal-level optimization, and motivate its extension to more challenging problems such as speech separation.

Building on our ICASSP 2026 work, SLM-SS~\cite{liSLMSSSpeechLanguage2026}, we present an extended SLM-SS framework that leverages speech language models (SLMs) together with quantized neural codecs to improve the intelligibility of separated speech. We adopt Encodec~\cite{encodec} to convert continuous waveforms into discrete multi-codebook token sequences, which makes speech separation amenable to language-modeling style sequence prediction. To represent multiple sources, we concatenate per-speaker token streams using Serialized Output Training (SOT)~\cite{sot}, following recent SOT-based ASR developments~\cite{sotasr,sotasr2}. The resulting sequence is generated with a transformer autoregressive (AR) encoder-decoder: the encoder extracts mixture representations, and the decoder aligns mixture features with target tokens via cross-attention. To further improve quality and reduce latency, we augment the AR backbone with a non-autoregressive (NAR) module that predicts higher-order codebooks conditioned on lower-order ones using independent, one-to-one token embeddings; we also propose a more efficient NAR variant to reduce training complexity. Compared with the ICASSP version, this paper provides a more complete formulation of hybrid AR/NAR modeling, introduces a per-talker NAR variant for efficient high-order token prediction, supports variable numbers of talkers, and includes more comprehensive codec comparisons as well as substantially expanded objective and subjective evaluations. Overall, the proposed framework offers a generative alternative to conventional SS approaches, improving speech intelligibility and yielding strong performance on downstream tasks.

Our contributions can be summarized as follows:
\begin{itemize}
\item We introduce \textbf{SLM-SS}, a generative speech separation framework that performs intelligibility-aware separation by predicting discrete neural-codec tokens with a speech language model.
\item We develop a hybrid AR/NAR generation scheme for multi-codebook token prediction, including a \textbf{per-talker NAR} variant that improves high-order codebook generation while keeping decoding efficient.
\item We compare two representative neural codecs (Encodec~\cite{encodec} and WavTokenizer~\cite{wavtokenizer}) and highlight the benefits of a \textbf{single-codebook} codec design, which simplifies training by removing the need for an additional non-autoregressive (NAR) decoding module.
\item We extend SLM-SS to support \textbf{a variable number of talkers}, improving flexibility in practical multi-speaker scenarios.
\item We conduct \textbf{extensive objective and subjective evaluations}, revealing complementary strengths of generative and discriminative separation approaches across linguistic, signal-level, and perceptual metrics.
\end{itemize}
\section{Related Works}
\label{sec:relatedworks}
\subsection{Speech Separation}
\begin{figure}[!ht]
    \centering
    \includegraphics[width=\linewidth]{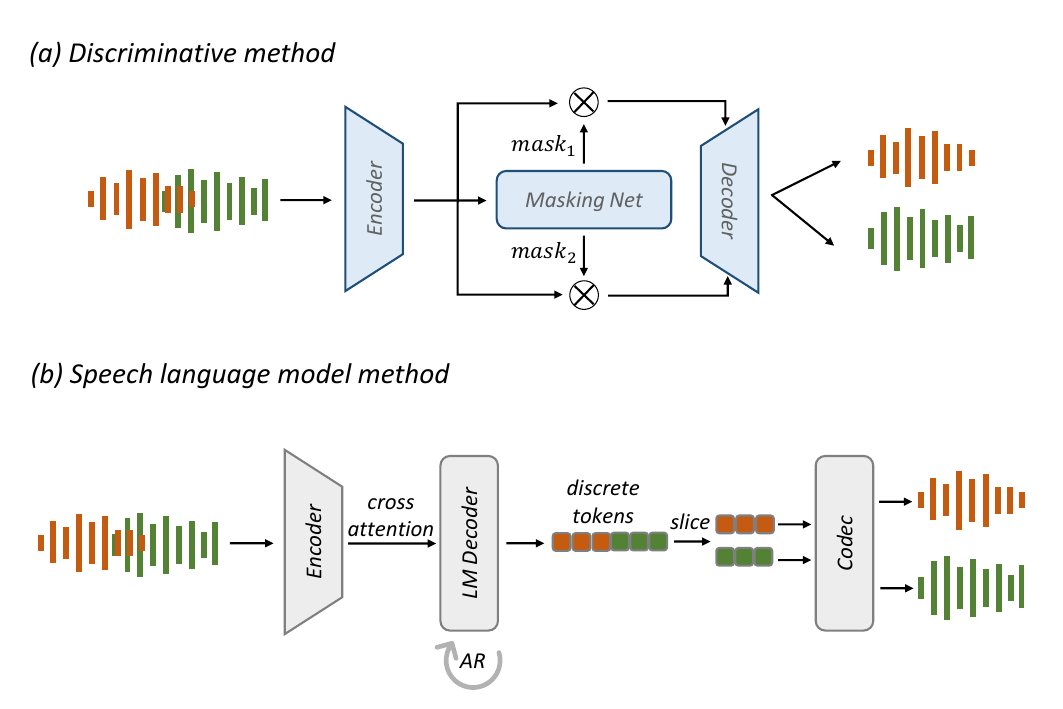}
    \caption{Structure of discriminative model and SLM model for speech separation.}
    \label{fig:comparison}

\end{figure}

Speech separation aims to isolate clean, single-talker speech from overlapping multi-talker mixtures that may also contain background noise. Most discriminative methods formulate separation as learning masking~\cite{ss-survey} or mapping~\cite{wangComplexSpectralMapping2020} functions under a fixed and known number of speakers (Fig.~\ref{fig:comparison}(a)). By estimating and applying different masks to intermediate audio representations, these approaches achieve the separation and \emph{time-domain} separation.

TF-domain separation performs separation on a spectrogram (e.g., STFT). It predicts masks or mapped spectra over \emph{time--frequency bins} and then reconstructs waveforms with an inverse transform, where phase handling is often an additional concern. Representative examples include Deep-CASA~\cite{deepcasa}, which performs frame-level spectral separation with permutation-invariant training (PIT)~\cite{pit,pit2} and uses Deep Clustering~\cite{deepclustering} for talker tracking, and BSRNN~\cite{bsrnn-luo2023}, which explicitly decomposes a mixture spectrogram into subbands and interleaves band-level and sequence-level modeling for improved performance.

In contrast, time-domain separation directly operates on waveforms or \emph{learned} time-domain latent representations without an explicit TF transform, thereby avoiding explicit phase reconstruction. TasNet~\cite{tasnet-luo2018} adopts an encoder--separator--decoder framework and performs separation on nonnegative latent representations, while Conv-TasNet~\cite{conv-tasnet} introduces a convolutional encoder to learn waveform representations optimized for separation and recovers sources via mask-based weighting. Subsequent works~\cite{luoDualPathRNNEfficient2020,chenDualPathTransformerNetwork2020,liSkimSkippingMemory2022,sepformer-cem2021} further improve time-domain modeling by capturing both short- and long-term dependencies with a multi-scale transformer-based architecture.

Beyond fixed-speaker settings, TUSS~\cite{tuss} and Flexio~\cite{flexio} explore multi-domain separation with a variable number of learnable prompts to specify target sources. SepEDA~\cite{sepeda}, SD-SepEDA~\cite{SD-SepEDA} and MUSE~\cite{MUSE} use triple-path processing blocks and attractor generation to support a variable number of talkers, while EEND-EDA~\cite{eendeda} integrates the Encoder-Decoder Attractor (EDA) module into an end-to-end speaker diarization (SD) system to handle variable-talker scenarios. More broadly, unified source separation and extraction has been explored in USE~\cite{wangUSEUnifiedModel2026} and Promptsep~\cite{promptsep}, indicating a trend toward general and scalable separation frameworks that can dynamically adjust their behavior across multiple separation tasks.

\subsection{Language Model in Speech Processing}
Recent studies have demonstrated that combining SLMs with discrete feature representations is effective across many speech processing tasks. For text-to-speech synthesis (TTS) task, VALL-E~\cite{valle} and Seed-TTS~\cite{seedtts-anastassiou2024} merge discrete audio tokens from neural codecs with text tokens to build a unified vocabulary. They then use an LM to generate stylistically consistent speech via autoregressive continuation conditioned on text and audio prompts. For speech recognition, Seed-ASR~\cite{seedasr-bai2024} and FireRedASR~\cite{fireredasr-xu2025} incorporate LLMs to achieve state-of-the-art ASR performance. For target speaker extraction (TSE), TSELM uses discrete WavLM outputs~\cite{wavlm} to encode target-speaker information and reconstruct speech from tokens~\cite{tangTSELMTargetSpeaker2026}. UniSep models SoundStream-derived discrete sequences~\cite{soundstream} and uses prompt audio for conditional target sound separation~\cite{unisep}. For general audio separation, OpenSep~\cite{opensep} leverages descriptive captions produced by an external audio captioning model to capture source-specific characteristics. SepALM targets noise-related issues by using an LM for end-to-end denoising of separated speech~\cite{sepalm}. SELM~\cite{SELM} extracts WavLM-based discrete representations and models them with a speech language model for speech enhancement (SE). High-fidelity token-based speech enhancement further extends this line of work~\cite{lanzendorferHighFidelitySpeechEnhancement2026}. TokenSplit introduces a multitask masking framework that jointly performs speech separation, ASR, and speech synthesis, optionally conditioned on transcripts~\cite{tokensplit}.

Despite these advances, most SLM-based approaches focus on single-stream modeling or conditional extraction, rather than structured multi-talker separation with explicit sequence-level constraints. As illustrated in Fig.~\ref{fig:comparison}(b), we instead model multi-talker speech separation end-to-end, without conditional prompts or external priors. Our preliminary ICASSP study (SLM-SS~\cite{liSLMSSSpeechLanguage2026}) validated the core idea of casting separation as discrete multi-codebook sequence generation. This journal version substantially extends SLM-SS in both modeling and evaluation:
\begin{itemize}
\item \textbf{Improved decoding/training for better quality--efficiency trade-off.} Beyond the VALL-E-inspired training recipe used in ICASSP, we explicitly consider speech quality and training cost, and propose a per-talker non-autoregressive (NAR) variant to enable efficient high-order decoding.
\item \textbf{Codec comparison and simplified training.} While the ICASSP work used only Encodec~\cite{encodec}, this journal version systematically compares different discrete tokenizers/codecs and analyzes their impact on separation performance. In particular, by adopting WavTokenizer~\cite{wavtokenizer}, which produces a single-codebook token sequence, we no longer require the NAR module, substantially simplifying the overall training pipeline.
\item \textbf{Variable-talker experiments.} We evaluate scalability and generalization to mixtures with varying numbers of speakers.
\item \textbf{More detailed ablations and comprehensive evaluation.} We conduct finer-grained ablations to quantify the effects of different autoregressive decoding strategies and other key design choices. We also add both objective metrics and subjective listening tests, and report results across multiple codecs.
\end{itemize}

\section{Speech Language Model for Speech Separation}
\label{sec:method}
\subsection{Overview}

\begin{figure*}[!ht]
    \centering
    \includegraphics[width=1\textwidth]{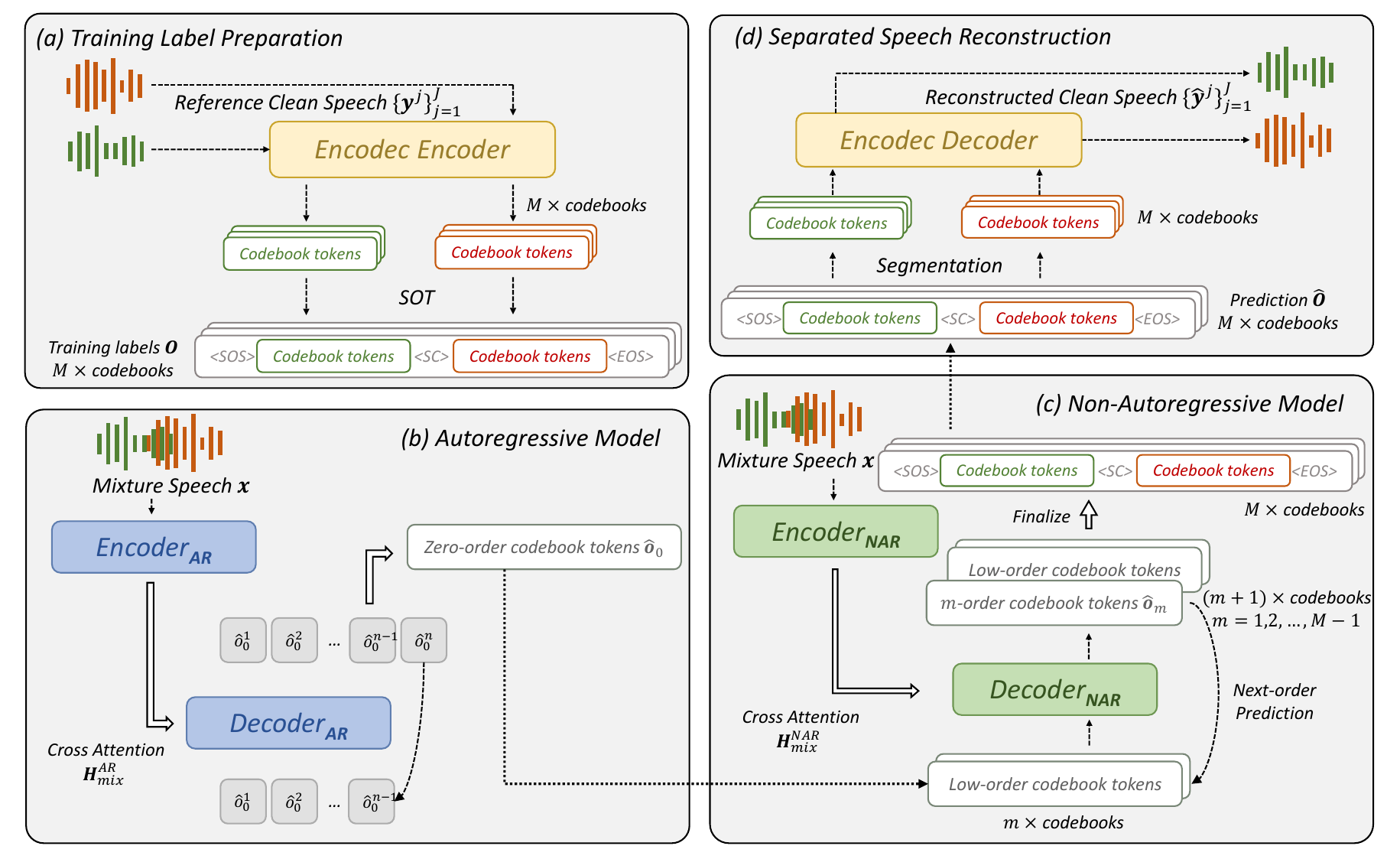}
    \caption{Overview of SLM-SS. (a) Encodec quantizes single-talker audio into multi-codebook sequences, then merges them using SOT for the training label preparation. (b) The AR model predicts the zero-order sequence. (c) NAR model sequentially predicts higher-order codebook sequences given lower-order ones. (d) Segment the SOT sequences into single-person sequences, then decoded into audio, respectively.}
    \label{fig:pipeline}

\end{figure*}

Fig.~\ref{fig:pipeline} summarizes the overall pipeline of SLM-SS. As shown in Fig.~\ref{fig:pipeline}(a), we first construct discrete supervision by quantizing each reference single-talker waveform with Encodec~\cite{encodec} into a multi-codebook token representation. Encodec employs residual vector quantization: the zero-order codebook carries the most salient spectral-temporal structure, while higher-order codebooks encode the remaining residual details. We then serialize tokens from different talkers into a single target sequence using serialized output training (SOT)~\cite{sot}.
As shown in Fig.~\ref{fig:pipeline}(b), we learn an attention-based encoder-decoder (AED) model to predict the zero-order SOT sequence in an autoregressive (AR) manner, using cross-attention over mixture features to align the mixture with the target tokens. Then, a non-autoregressive (NAR) model (Fig.~\ref{fig:pipeline}(c)) with the same backbone is further used to predict higher-order codebooks conditioned on all lower-order sequences, serving as a refinement stage. This AR-NAR decomposition follows the spirit of VALL-E~\cite{valle}: AR modeling focuses on the most informative codebook, while NAR modeling completes the higher-order residual codebooks.
Finally, as shown in Fig.~\ref{fig:pipeline}(d), we split the serialized sequence into per-talker streams using the special speaker change (\texttt{<SC>}) tokens, and decode the resulting multi-codebook tokens with the Encodec decoder to obtain the separated single-talker waveforms.

\subsection{Training Label Preparation}

We formulate speech separation as discrete sequence generation. We use Encodec~\cite{encodec} to quantize each single-talker waveform into a $M$-order ($M=32$ in Encodec) codebook representation, where each $m$-th ($m = 0, 1,  \dots, M-1$) order codebook has a base vocabulary $\mathcal{V}_{m}^{'}$ with size $|\mathcal{V}_{m}^{'}| = 1024$. For serialized output training (SOT), we further augment each codebook vocabulary with three special symbols (\texttt{<SOS>}, \texttt{<SC>}, \texttt{<EOS>}), and denote the resulting vocabulary by:
\begin{equation}
    \mathcal{V}_m = \mathcal{V}_m^{'} \cup \{\texttt{<SOS>},\texttt{<SC>},\texttt{<EOS>}\}.
\end{equation}
Note that different-order codebooks all have the same base vocabulary size $|\mathcal{V}_m| = 1027$ (hence the same token-ID range); however, token IDs are defined independently for each order $m$, so the same numerical ID across orders is not comparable and carries no shared meaning. In contrast, the special symbols maintain consistent semantic meanings across all $M$ codebooks, enabling unambiguous representation of structural information such as talker transitions and sequence boundaries.

Our training set is constructed via data simulation: each multi-talker mixture is synthetically generated by mixing multiple clean single-talker utterances, so the underlying clean sources are known and can be directly used as supervision. As Fig.~\ref{fig:pipeline}(a) shows, for each simulated mixture, we independently encode every constituent clean utterance with Encodec to obtain its multi-codebook token sequences. To prepare a single label sequence for multi-talker training, we adopt serialized output training (SOT)~\cite{sot} to serialize talker-wise token streams into one target $\mathbf{O}$:
\begin{equation}
    \mathbf{O} = [\mathbf{o}_0, \mathbf{o}_1, \dots, \mathbf{o}_{M-1}],
\end{equation}
where $\mathbf{o}_m$ is the $m$-th order codebook training label.
For the $m$-th order codebook sequence $\mathbf{o}_m$, we prepend \texttt{<SOS>}, concatenate the token sequences of different talkers in utterance first-in-first-out order, insert \texttt{<SC>} between adjacent talkers, and append \texttt{<EOS>}:
\begin{equation}
\label{eq:om}
\begin{aligned}
\mathbf{o}_m &= \Bigl[\texttt{<SOS>},\, r_{1,m}^{1},\dots,r_{K^{1},m}^{1},\,\texttt{<SC>},\\
&\qquad r_{1,m}^{2},\dots,r_{K^{2},m}^{2},\,\texttt{<SC>},\,\dots,\,\texttt{<SC>},\\
&\qquad r_{1,m}^{J},\dots,r_{K^{J},m}^{J},\,\texttt{<EOS>}\Bigr].
\end{aligned}
\end{equation}
where $r_{k,m}^j \in \mathcal{V}_m^{'}$ denotes the $k$-th token of the $j$-th talker in the $m$-th order codebook sequence, and $K^j$ represents the token sequence length corresponding to the $j$-th talker. 

\subsection{AR Model for Zero-order Codebook Sequence Decoding}
During training, the AED model is required to predict the zero-order codebook sequence, which carries the dominant spectral and temporal structure of the target speech. For the encoder, we adopt a pre-trained WavLM~\cite{wavlm} model and finetune it for speech separation. For the decoder, we follow the architectural design of the Whisper~\cite{whisper} decoder and train it from scratch for discrete token generation.

Given a multi-talker input waveform, the encoder extracts hidden representations from all WavLM layers.
We then use a linear fusion layer followed by layer normalization to aggregate them into a unified acoustic representation $\mathbf{H}_{\mathrm{mix}}^{\mathrm{AR}}$. The AR model mainly learns the zero-order codebook at $m=0$.  
For the sake of convenience in description, we rewrite the training label $\mathbf{o}_0$ in Eq.\ref{eq:om} into the following form when $ m = 0$:
\begin{equation}
    \mathbf{o}_0 = [o_0^1,\dots,o_0^{N}],
\end{equation}
where $o_0^n$ is the $n$-th token in the training label $\mathbf{o}_0$ and $N$ is the length of $\mathbf{o}_0$. During training, conditioned on $\mathbf{H}_{\mathrm{mix}}^{\mathrm{AR}}$ and the teacher-forced history $[o_0^1,\dots,o_0^{n-1}]$, the autoregressive decoder is trained to predict the next token $\hat{o}_0^{n}$:

\begin{equation}
    \mathbf{q}_0^{n} = \text{Decoder}([o_0^1,\dots,o_0^{n-1}],\mathbf{H}_{\mathrm{mix}}^{\mathrm{AR}}).
\end{equation}
where $\mathbf{q}_0^{n}\in \mathbb{R}^{|\mathcal{V}_0|}$ is the probability distribution of the $n$-th token. During inference, we decode a predicted token $\hat{o}_0^{n}$ from $\mathbf{q}_0^{n}$, and the decoder conditions on its own previously generated history $[\hat{o}_0^1,\dots,\hat{o}_0^{n-1}]$. In this way, the AED model learns to map a multi-talker mixture directly to the zero-order SOT sequence, thereby capturing both talker serialization and coarse speech content in a single autoregressive generation process. We optimize the AR decoder with the standard cross-entropy loss:
\begin{equation}
\mathcal{L}_{\mathrm{AR}} = -\sum_{n=1}^{N} \log {q}_0^{n}\!\left(o_0^{n}\right),
\end{equation}
where one-hot labels are used, and ${q}_0^{n}\!\left(o_0^{n}\right)$ denotes the predicted probability assigned to the ground-truth token $o_0^{n}$ (i.e., the $o_0^{n}$-th entry of $\mathbf{q}_0^{n}$).

\subsection{NAR Model for High-order Codebook Sequences Decoding}
\begin{figure}[!ht]
    \centering
    \includegraphics[width=\linewidth]{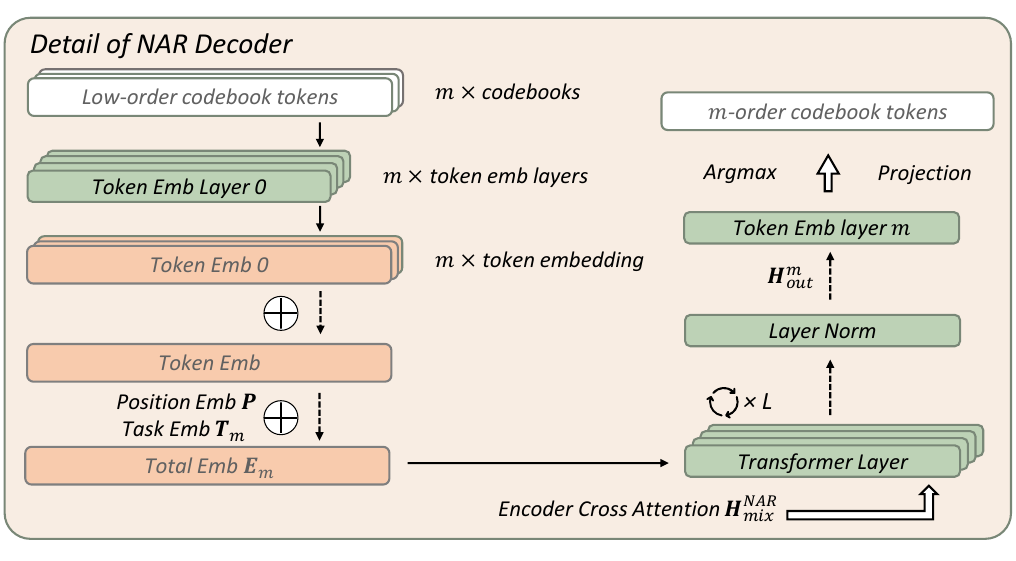}
    \caption{Detail of NAR Decoder. NAR decoder integrates all low-order sequence information by employing multiple independent token embeddings.}
    \label{fig:nar}

\end{figure}

Inspired by the AR+NAR decomposition in VALL-E~\cite{valle}, we adopt the same backbone as the AED model and remove the unidirectional attention mask to implement a non-autoregressive (NAR) decoder for high-order codebook prediction. Unlike VALL-E, whose NAR stage is purely conditioned on discrete codec tokens, our NAR model is equipped with an encoder and takes continuous mixture features as input, enabling refined prediction with richer acoustic conditioning. As illustrated in Fig. \ref{fig:nar}, to enable collaborative training across multiple codebook orders, we design several independent token embedding layers with one-to-one correspondence to different orders of codebook sequences, and share a common positional embedding, allowing different orders to be modeled in a coordinated manner over the entire SOT sequence. In addition, we introduce a task embedding $\mathbf{T}_m$ that is obtained by encoding the target codebook order index $m$ (with $m\in\{1,\dots,M-1\}$), explicitly specifying that the current task is to predict the $m$-th order codebook; this enables the model to effectively distinguish among different prediction tasks during both training and inference.

When predicting the $m$-th order codebook sequence, information from all lower-order codebook sequences must be jointly considered. To this end, all lower-order sequences are first embedded independently and then summed together. The positional embedding and the corresponding task embedding are subsequently incorporated to form the total embedding $\mathbf{E}_m$, which serves as the input to the NAR model, namely:

\begin{equation}
\mathbf{E}_m = \Bigg( \sum_{\ell=0}^{m-1} \mathrm{Emb}\big(\mathbf{o}_{\ell};\mathbf{W}_{\ell}\big) \Bigg) 
      + \mathbf{P} + \mathbf{T}_m ,
\end{equation}
where $\mathrm{Emb}\big(\mathbf{o}_{\ell}; \mathbf{W}_{\ell} \big)$, $\ell \in [0, m)$, denotes the token embedding of the $\ell$-th order codebook sequence $\mathbf{o}_{\ell}$ with the $\ell$-th token embedding layer $\mathbf{W}_{\ell}$, $\mathbf{P}$ represents the positional embedding, and $\mathbf{T}_m$ is a learned task embedding obtained from an encoding of the order index $m\in\{1,\dots,M-1\}$, indicating that the decoder should predict the $m$-th order codebook. The resulting embedding $\mathbf{E}_m$ is then fed into a stack of Transformer layers for deep contextual modeling. Finally, the output representations $\mathbf{H}_{\mathrm{out}}^m$ are projected onto the token embedding space of the $m$-th order codebook $\mathbf{W}_m$, yielding the probability distribution $\mathbf{q}_m$ over tokens in the $m$-th order vocabulary $\mathcal{V}_m$:

\begin{equation}
\begin{aligned}
\mathbf{H}_{\mathrm{out}}^m &= \mathrm{DecoderTransformer}(\mathbf{E}_m), \\
\mathbf{q}_m &= \mathrm{Softmax}\!\left(\mathbf{H}_{\mathrm{out}}^m \mathbf{W}_m^\top \right).
\end{aligned}
\end{equation}
To train the NAR model, we optimize it with the cross-entropy loss over all high-order codebook sequences. Specifically, for each $m \in \{1,\dots,M-1\}$, we denote the training label as $\mathbf{o}_m = [o_m^1,\dots,o_m^{N}]$, and let $\mathbf{q}_m^{n}\in\mathbb{R}^{|\mathcal{V}_m|}$ be the predicted distribution at position $n$ (i.e., the $n$-th row of $\mathbf{q}_m$). The NAR loss can be defined as:
\begin{equation}
\label{eq:loss_nar}
\mathcal{L}_{\mathrm{NAR}} = -\sum_{m=1}^{M-1}\sum_{n=1}^{N} \log {q}_m^{n}\!\left(o_m^{n}\right).
\end{equation}
It is noted that outer summation over $m$ in Eq.~\ref{eq:loss_nar} represents our desired objective of optimizing all high-order codebooks jointly. However, directly computing the full sum over $m\in\{1,\dots,M-1\}$ for every training sample is memory-intensive, since each $m$-th prediction conditions on all lower-order sequences. Therefore, in practice we randomly sample a specific $m\sim\mathrm{Uniform}(\{1,\dots,M-1\})$ for each training sample and optimize only the corresponding term, which serves as a stochastic approximation of Eq.~\ref{eq:loss_nar}.
Meanwhile, because the NAR model must integrate information from all lower-order codebook sequences to predict higher-order ones, its computational cost and training challenges (e.g., convergence time and sensitivity to hyperparameters) increase with the total order of the codebooks. In this work, we predict codebooks up to the 8-th order out of 32 codebooks in Encodec (i.e., $M=8$); as we will show in Sec.~\ref{sec:experiment}, this truncation has a limited impact on performance.

\subsection{Inference Procedure and Speech Reconstruction}

During inference, the trained AED model first receives the multi-talker mixture and autoregressively generates the predicted zero-order codebook sequence $\hat{\mathbf{o}}_0$ in SOT format. Here, we use $\hat{\mathbf{o}}_m$ to denote the decoded (predicted) $m$-th order codebook sequence, distinguishing it from the ground-truth training label $\mathbf{o}_m$. The predicted zero-order sequence already determines the talker ordering and the segmentation boundaries through the predicted \texttt{<SC>} symbols. The generated $\hat{\mathbf{o}}_0$ is then used in two ways. First, it provides the low-order input required by the NAR model to predict higher-order codebooks, as described in the next subsection. Second, after all $m$ codebook orders have been obtained, the complete SOT sequence is segmented according to the \texttt{<SC>} symbols to recover the single-person token sequences.

Each segmented single-person predicted multi-codebook sequence is finally passed to the Encodec decoder to reconstruct the separated waveform. Because the number of talkers is implicitly represented by the number of predicted separator symbols \texttt{<SC>}, the proposed framework can naturally handle mixtures with an unknown talker count at inference time, without requiring an additional talker-count estimator or prior assumptions on the number of talkers.

\subsection{Per-Talker NAR: Simplified NAR Decoding Strategy}
As discussed earlier, training joint multi-talker NAR models poses significant challenges due to quadratic growth in computational cost and convergence difficulty as the number of codebooks increases. However, we observe that the most critical information for speech separation is already determined by the zero-order codebook sequence produced by the AR model. Since the zero-order codebook contains dominant spectral and temporal structure, requiring the NAR model to jointly refine codebooks across all speakers may introduce unnecessary complexity.

Motivated by this observation, we propose an alternative per-talker NAR approach.
Instead of jointly refining all speakers' codebooks in the SOT format, we first segment the multi-talker zero-order sequence $\hat{\mathbf{o}}_{0}$ using the predicted separator symbols \texttt{<SC>}, then independently apply the NAR model to each single-talker segment.
This approach leverages the mixture representation together with each talker's accumulated low-order tokens to predict the corresponding higher-order codebooks. By design, the per-talker NAR avoids modeling speaker transitions at higher orders, substantially reducing task complexity and computational burden. The full inference procedure and notation are summarized in Algorithm \ref{alg:singlenar}.

\begin{algorithm}[H]
\caption{Inference with Per-talker NAR Scheme}
\label{alg:singlenar}
\renewcommand{\algorithmicrequire}{\textbf{Input:}}
\renewcommand{\algorithmicensure}{\textbf{Output:}}
\begin{algorithmic}[1]
\REQUIRE Mixture waveform $\mathbf{x}$, number of codebook orders $M$.
\ENSURE Reconstructed waveform $\{\hat{\mathbf{y}}^{j}\}_{j=1}^{J}$ of each talker .
\item[\textbf{Notation:}] {\footnotesize\hfill\\
\hspace*{\algorithmicindent}$\hat{\mathbf{o}}_{m}$: predicted $m$-th order codebook sequence.\\
\hspace*{\algorithmicindent}$J$: talker count inferred from $\hat{\mathbf{o}}_{0}$ by $\texttt{<SC>}$.\\
\hspace*{\algorithmicindent}$\hat{\mathbf{o}}_{0}^{j}$: $j$-th segment of $\hat{\mathbf{o}}_{0}$ split by $\texttt{<SC>}$.\\
\hspace*{\algorithmicindent}$\mathbf{H}_{\mathrm{mix}}^{\mathrm{AR}}/\mathbf{H}_{\mathrm{mix}}^{\mathrm{NAR}}$: mixture features from AR/NAR encoders.\\
\hspace*{\algorithmicindent}$\mathbf{Z}^{j}$: accumulated tokens for talker $j$ (init $\hat{\mathbf{o}}_{0}^{j}$; append $\hat{\mathbf{o}}_{m}^{j}$ for $m\ge1$).\\}
\STATE {\textsc{INFERENCE FUNCTION}}$(\mathbf{x}, M)$
\STATE \hspace{0.5cm}$ \mathbf{H}_{\mathrm{mix}}^{\mathrm{AR}} \gets \mathrm{Encoder}_{\mathrm{AR}}(\mathbf{x}) $
\STATE \hspace{0.5cm}$ \hat{\mathbf{o}}_{0} \gets \mathrm{Decoder}_{\mathrm{AR}}(\mathbf{H}_{\mathrm{mix}}^{\mathrm{AR}}) $
\STATE \hspace{0.5cm}$ \{\hat{\mathbf{o}}_{0}^{j}\}_{j=1}^{J} \gets \mathrm{Split}(\hat{\mathbf{o}}_{0}, \texttt{<SC>}) $
\STATE
\STATE \hspace{0.5cm}$ \mathbf{H}_{\mathrm{mix}}^{\mathrm{NAR}} \gets \mathrm{Encoder}_{\mathrm{NAR}}(\mathbf{x}) $
\STATE
\STATE \hspace{0.5cm}\textbf{for} $j = 1, \ldots, J$

\STATE \hspace{1cm}$ \mathbf{Z}^{j} \gets \hat{\mathbf{o}}_{0}^{j} $
\STATE
\STATE \hspace{1cm}\textbf{for} $m = 1, \ldots, M-1$
\STATE \hspace{1.5cm}$ \hat{\mathbf{o}}_{m}^{j} \gets \mathrm{Decoder}_{\mathrm{NAR}}(\mathbf{H}_{\mathrm{mix}}^{\mathrm{NAR}}, \mathbf{Z}^{j}) $
\STATE \hspace{1.5cm}$ \mathbf{Z}^{j} \gets \mathrm{Append}(\mathbf{Z}^{j}, \hat{\mathbf{o}}_{m}^{j}) $
\STATE
\STATE \hspace{1cm}$ \hat{\mathbf{y}}^{j} \gets \mathrm{Decoder}_{\mathrm{Codec}}(\mathbf{Z}^{j}) $
\STATE
\STATE \hspace{0.5cm}\textbf{return} $\{\hat{\mathbf{y}}^{j}\}_{j=1}^{J}$
\end{algorithmic}
\end{algorithm}

\subsection{Fully AR Scheme with a Single-Order Codec}
Although Encodec provides strong reconstruction quality, its multi-order residual vector quantization design requires the additional NAR stage to recover higher-order codebooks, which increases both training complexity and inference cost. To examine whether the separation framework can be further simplified, we additionally consider a fully autoregressive variant built on a single-order neural audio codec.

Specifically, we replace Encodec with WavTokenizer~\cite{wavtokenizer}, which represents each utterance using only one codebook sequence while maintaining high-fidelity reconstruction. Under this setting, the AED model directly predicts the complete target token sequence, and no higher-order codebook refinement is required. As a result, the overall system no longer needs the NAR module, and the separation pipeline is reduced to three steps: single-order token generation, SOT-based sequence segmentation, and codec decoding.

This variant serves two purposes. First, it provides a computationally simpler alternative to the Encodec-based AR-NAR framework. Second, it allows us to analyze how codec design, especially the choice between multi-order and single-order tokenization, affects downstream speech separation performance. Experimental comparisons between this scheme and the Encodec-based methods are reported in Sec. \ref{sec:experiment}.

\section{Experiment}

\label{sec:experiment}
\subsection{Experiment Setup}
\textbf{Model:}
The encoder is initialized with pre-trained WavLM-large weights~\cite{wavlm}, while the decoder adopts the dimensional configuration of Whisper-medium~\cite{whisper}, with an adjusted vocabulary size and a reduced depth of 16 Transformer layers to save computational resources. The resulting model comprises approximately 600M parameters in total. Training is performed for 30 epochs using an initial learning rate of $5\times10^{-5}$, together with cosine annealing decay and linear warm-up during the first three epochs.

For AR decoding, we employ beam search with $beamsize=20$, and apply blank suppression as well as N-gram blocking to effectively prevent empty predictions and infinite repetition during inference. For the NAR model and its variant, due to training difficulties arising from multi-order dependency modeling, we adopt a staged training strategy. Specifically, the model is first trained on the first-order codebook prediction task, where the zero-order codebook sequence generated by the AR model is used to predict the first-order codebook sequence. During this stage, embedding layers corresponding to higher-order codebook sequences are excluded from computation. Subsequently, the resulting model weights are retrained on all non-zero codebook prediction tasks using identical training configurations, enabling stable convergence across multiple codebook orders.

\textbf{Dataset \& Baseline:}
Experiments are conducted on LibriMix~\cite{librimix}, max and clean version with 16kHz sampling rate, using the 100h and 360h subsets for training and the test subset for evaluation. We follow the original LibriMix data simulation pipeline for mixture generation, and additionally apply preprocessing to clean speech before Encodec quantization: VAD-based segmentation is used to remove silent regions. This extra step has two purposes. First, it reduces unnecessary silent tokens in codebook sequences and thus lowers computational overhead. Second, and more importantly for SOT modeling, it avoids variable-length silence spans that can shift token positions and even invert the effective temporal order of speakers in the generated SOT sequence. Since Encodec-based encoding and decoding, as well as partial codebook truncation, may introduce distortions that affect signal quality, we explicitly examine the relationship between the original audio and its Encodec-processed counterparts.

Although the clean waveform is the nominal reference, directly modeling all 32 Encodec codebooks is impractical in our AR--NAR setup: when predicting higher-order codebooks, the NAR model must condition on all lower-order codebooks, which substantially increases optimization difficulty and computational cost as the order grows. Following the coarse-to-fine codec language modeling paradigm used in VALL-E~\cite{valle}, we focus on the first 8 codebooks, which retain most intelligibility-relevant and linguistic information, while treating higher-order codebooks mainly as residual detail refinement. 
Therefore, in our setting, the practical upper bound (UB) is not the original waveform groundruth (GT), or GT-Encodec32 that has undergone a full encoding-decoding reconstruction by Encodec, but GT-Encodec8 reconstructed from the same 8 codebooks used for training. 
For comparison, we adopt BSRNN~\cite{bsrnn-luo2023}, SepFormer~\cite{sepformer-cem2021}, TF-LocoFormer~\cite{tflocoformer}, and MossFormer2~\cite{mossformer2} as baseline methods.

\textbf{Evaluation Metrics:}
We conduct both subjective (only for 6 selected systems) and objective tests to evaluate the quality of the SLM's speech separation. 
For the objective test, we report speaker similarity, word error rate (WER) by pretrained Whisper~\cite{whisper} base model, Levenshtein Phoneme Similarity (LPS)~\cite{lps}, and SpeechBERTScore (SBS)~\cite{speechbertscore}. In terms of signal quality, we recorded the PESQ~\cite{pesq} (wb) and tested the non-intrusive perceptual metrics UTMOS~\cite{utmos} and DNSMOS~\cite{dnsmos}. In addition, to further analyze Encodec’s and WavTokenizer's internal consistency, we compute the token error rate (TER), which is obtained between the predicted zero-order codebook sequence and the ground truth zero-order codebook. It is important to note that the TERs of the two neural codecs are not comparable. For the comparison models, the main framework’s Encodec is used uniformly for TER computation. For fair evaluation, discriminative baseline outputs are truncated according to the source-active timestamps retained during data simulation, so that all systems are evaluated over consistent active-speech regions.
For the subjective test, we conduct a blind multi-stimulus listening test of perceived speech quality with 30 volunteers of different ages, genders, and occupations on 16 single-source samples from 8 mixtures.
For each sample, listeners hear all 6 anonymized versions in randomized order before rating the overall perceived quality of each. These comprise 3 SLM-SS configurations, 2 strong discriminative baselines, and the original recording. We limit each trial to 6 conditions because comparing too many versions would make relative quality judgments more difficult.

\subsection{Results of the Objective Metrics}

Our experimental results are summarized in Table \ref{table:result}.
As shown in the table, SLM-SS achieves a WER of $7.24\%$, substantially outperforming BSRNN ($20.02\%$) and SepFormer ($17.77\%$), and also surpassing MossFormer2 ($8.78\%$), while TF-LocoFormer achieves the lowest WER ($5.88\%$). A similar metric-dependent pattern is observed for LPS and SBS: SLM-SS outperforms BSRNN and SepFormer but does not consistently surpass the more recent discriminative baselines. Its speaker similarity is slightly higher than that of TF-LocoFormer but lower than those of BSRNN, SepFormer, and MossFormer2, indicating competitive rather than uniformly superior speaker-identity preservation. Unlike discriminative models with sample-aligned outputs, autoregressive SLMs do not explicitly constrain the generated sequence length to the reference duration. Although wideband PESQ performs delay alignment and zero-padding, it cannot compensate for global time-scale differences; thus, PESQ may also reflect residual temporal drift and duration mismatch rather than waveform-reconstruction quality alone.

Introducing the per-talker NAR module consistently improves all reported metrics over the original SLM-SS. Its WER and TER approach those of GT-Encodec8, suggesting that decoupling NAR reconstruction from the SOT separation task improves reconstruction quality. The WavTokenizer-based AR-only model retains comparable speaker similarity and SBS, with moderately higher WER and lower PESQ, but improves UTMOS and DNSMOS over the Encodec-based models. These results suggest that WavTokenizer favors perception-oriented quality without uniformly improving linguistic accuracy or intrusive signal fidelity, reflecting different optimization objectives of the two neural codecs.


To further contextualize the objective results in Table \ref{table:result}, we analyze how the distortion introduced by neural codecs affects these metrics. Even when restoring the full 32-order codebooks of Encodec, the reconstructed audio still exhibits noticeable information loss, which becomes more pronounced when only the first 8 orders are retained. This irreversible distortion, introduced by feature discretization, results in slight degradation across objective metrics when compared to the original audio. However, such degradation is less perceptible in subjective listening tests, thereby supporting our hypothesis that discretization-induced artifacts do not always correlate with human auditory perception. A similar effect is reflected in TER: audio reconstructed from the full 32-order codebooks still produces token errors after re-encoding, confirming the persistence of discretization-related distortion. When only the predicted 8-order codebooks are decoded before re-encoding, SLM-SS exhibits stronger mismatches; nevertheless, the degradation remains less severe than that observed with SepFormer and BSRNN, benefiting from the Encodec-centered representation framework.

\begin{table*}[!htb]
    \caption{Overall comparison of SLM-SS with existing approaches.}
    \centering
    \begin{tabularx}{\textwidth}{lYYYYYYYYY}
        \toprule

        & \multicolumn{5}{c}{Speaker \& Linguistic Consistency} 
        & \multicolumn{1}{c}{Signal}
        & \multicolumn{2}{c}{Non-intrusive}
        & \multicolumn{1}{c}{\textbf{Subjective}} 
        \\
        \cmidrule(lr){2-6}  \cmidrule(lr){7-7}
        \cmidrule(lr){8-9}
        \textbf{Method}  & \textbf{Spk Sim.$\uparrow$} &  \textbf{WER}$\downarrow$& \textbf{TER}$\downarrow$ & \textbf{LPS}$\uparrow$ & \textbf{SBS}$\uparrow$ & 
        \textbf{PESQ}$\uparrow$ &
        \textbf{UTMOS}$\uparrow$ &
        \textbf{DNSMOS}$\uparrow$ &
        \textbf{Score}$\uparrow$\\
        \midrule
       \textbf{GT} & - & 5.19 & - & 1.000 & 1.000 & 4.64 & 4.09 & 3.59 & 78.41$\pm$4.08 \\
       \textbf{GT-Encodec32} &  93.5 & 6.03 & 24.7 & 0.975 & 0.957 & 3.14 & 3.18 & 3.50 & - \\
       \textbf{GT-Encodec8 (UB)} &  92.8 & 6.31 & 39.0 & 0.970 & 0.944 & 2.62 & 2.95 & 3.38 & -\\
       \textbf{GT-WavTokenizer} &  92.7 & 7.58 & $^*$47.3 & 0.964 & 0.939 & 2.33 & 4.01 & 3.67 & -\\
\cdashline{1-10}\noalign{\vskip\belowrulesep}
       \textbf{BSRNN} & 92.9 & 20.02 & 65.1 & 0.916 & 0.903 & 2.37 & 2.96 & 3.02 & -\\
       \textbf{SepFormer} & \textbf{93.4} & 17.77 & 70.0 & 0.949 & 0.910 & 2.78 & 3.21 & 3.15 & -\\
       \textbf{TF-LocoFormer} & 91.6 & \textbf{5.88} & 68.3 & \textbf{0.973} & 0.919 & \textbf{3.51} & 3.79 & 3.56 & 79.23$\pm$3.84\\
       \textbf{MossFormer2} & 92.2 & 8.78 & 62.6 & 0.944 & 0.916 & 2.95 & 3.27 & 3.24 & 71.59$\pm$4.56\\
        \midrule
       \textbf{SLM-SS} & 91.7 & 7.24 & 45.8 & 0.954 & 0.913 & 2.17 & 2.64 & 3.21 & 74.58$\pm$4.21\\
      \hspace{1em}\textbf{+ Per-talker NAR}
        & 92.3 & 6.92 & \textbf{41.7} & 0.959 & \textbf{0.925}
        & 2.30 & 2.88 & 3.36 & 76.83$\pm$4.22\\
       \hspace{1em}\textbf{+ WavTokenizer}
        & 92.3 & 8.97 & $^{*}$66.8 & 0.923 & 0.921
        & 2.06 & \textbf{3.95} & \textbf{3.65} & \textbf{81.41$\pm$3.69}\\
        \bottomrule
    \end{tabularx}
    \label{table:result}
    {\footnotesize
\raggedright
TER values marked with $^*$ are computed using WavTokenizer. 
They are not directly comparable with others using Encodec. Subjective scores are reported as mean $\pm$ 95\% confidence interval. We limit each subjective listening trial to 6 conditions because comparing too many versions would make quality judgments difficult for listeners.
\par}
\end{table*}

\subsection{Analysis of Subjective Listening Results}
Overall, both SLM-SS and the baseline methods inevitably introduce certain distortions and mismatches. Meanwhile, different evaluation metrics focus on distinct aspects of speech quality, which may lead to inconsistent trends across metrics. To more accurately reflect their practical impact, greater emphasis should therefore be placed on subjective listening tests. From this perspective, the audio reconstructed by the per-talker NAR SLM-SS remains largely comparable to the original recording and falls between the two advanced discriminative models. Furthermore, thanks to WavTokenizer’s targeted optimizations for signal quality and human auditory perception, the WavTokenizer-based approach achieved the highest mean subjective score. Pairwise Wilcoxon signed-rank tests further show that the improvement of SLM-SS-WT over the other evaluated systems is statistically significant after Holm correction. These results indicate that codec-induced distortions do not necessarily translate into a comparable degradation in perceived speech quality.

\begin{figure*}[!ht]
    \centering
    \includegraphics[width=1\textwidth]{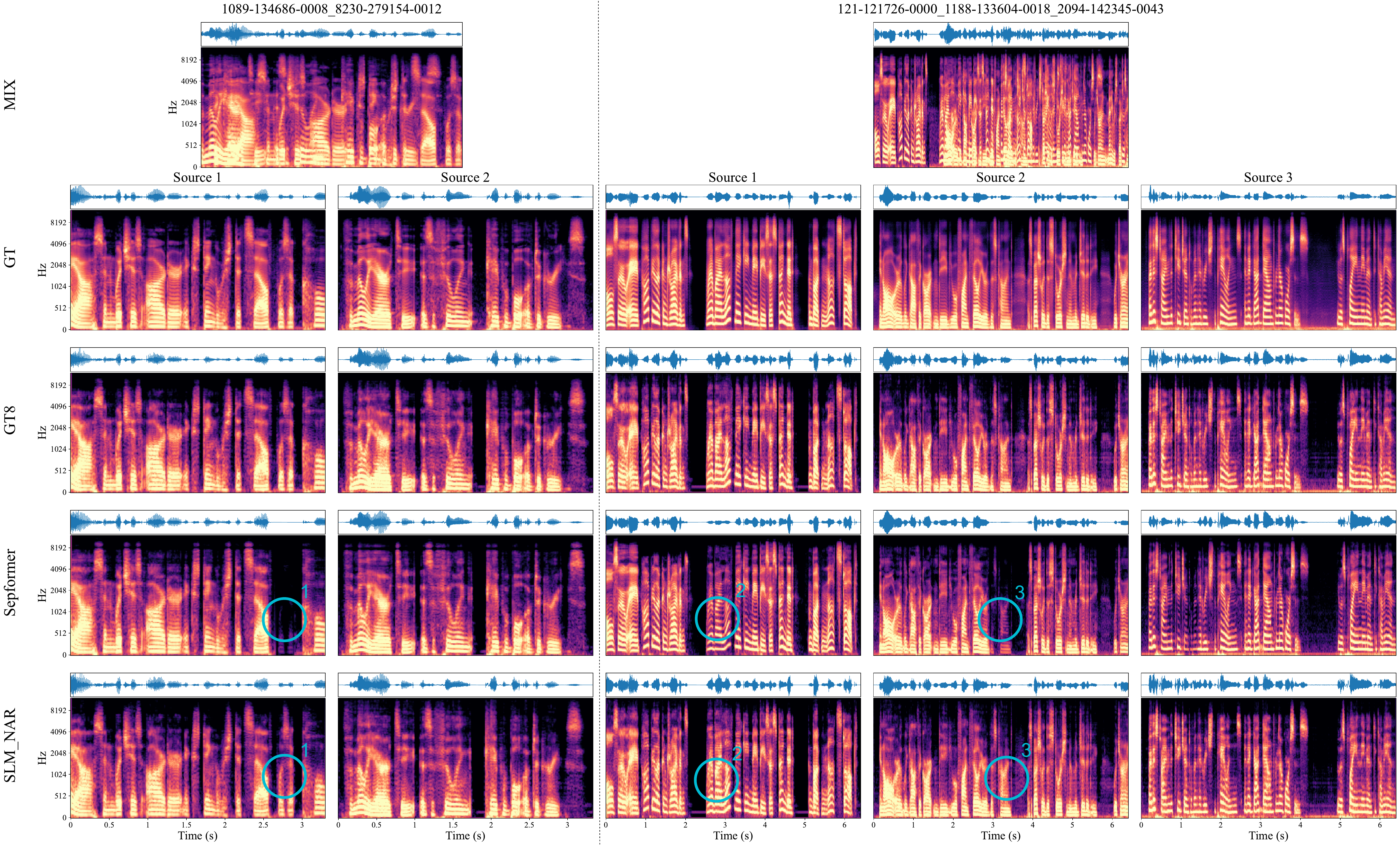}
    \caption{Comparison of time series and mel-frequency spectrograms across different models. Truncating Encodec to the first 8 codebooks (GT8) removes high-frequency details and slightly degrades mel-spectrogram clarity while preserving formant structure, leading SLM-SS to exhibit similar spectral characteristics. In contrast, SepFormer retains finer spectral resolution by directly operating on waveforms, but may introduce over-suppression artifacts under challenging conditions such as heavy overlap or imbalanced talker energy, as marked areas 1, 2, and 3.}
    \label{fig:demo}

\end{figure*}

To more clearly compare the differences in the audio generated by the two paradigms, we selected a few samples for visualization, as shown in Fig. \ref{fig:demo}. It can be observed that truncating Encodec to only the first 8 codebooks (GT8) leads to the loss of high-frequency components, resulting in a degradation in the clarity of the mel-spectrogram. However, this compression does not cause significant formant smearing, and the harmonic structure remains largely distinguishable. Since the proposed SLM-SS framework adopts GT-Encodec8 (GT8) as an upper performance bound, its spectral characteristics are largely consistent with those of GT-Encodec8.

In contrast, SepFormer operates directly on the waveform domain and thus preserves more detailed spectral information, exhibiting less degradation in mel-spectrogram resolution. Nevertheless, under challenging conditions such as severe source overlap or highly imbalanced talker energy, SepFormer tends to introduce over-suppression artifacts, which may lead to the attenuation or loss of certain speech segments.

\subsection{Ablation Study on Temperature During Inference}
In this experiment, we investigate whether the proposed method requires temperature tuning during inference.
The motivation of this ablation is threefold. First, in autoregressive generation, temperature directly controls token uncertainty and therefore affects reconstruction quality. Second, it influences decoding stability: excessively low temperature often leads to over-confident but rigid token selection, while excessively high temperature can cause unstable token trajectories and error accumulation. Third, it is closely related to hallucination behavior in codec-token generation (e.g., speaker confusion, non-speech artifacts, or content drift). This concern is also consistent with prior neural codec language modeling practice, including VALL-E~\cite{valle}, where temperature is commonly tuned at inference to balance fidelity and diversity. Therefore, we conduct a systematic ablation over temperature to verify whether SLM-SS requires such tuning and to identify a robust operating point. As reported in Table~\ref{table:temp}, the proposed method achieves optimal results using the default temperature setting of 1.0, without requiring additional tuning.  This result suggests that our model can maintain good generation quality and stable decoding with minimal risk of temperature-induced hallucination, without additional inference-time tuning.

\begin{table}[!htp]
    \caption{Performance on different AR temperatures.}
    \centering
    \begin{tabular}{cccccc}
        \toprule
        \textbf{Temperature} & \textbf{Spk Sim.$\uparrow$} & \textbf{WER}$\downarrow$ & \textbf{TER}$\downarrow$ & \textbf{LPS}$\uparrow$ & \textbf{SBS}$\uparrow$\\
        \midrule
        0.5 & 38.9 & 49.08 & 69.3 & 0.581 & 0.695 \\
        0.9  & 73.1 & 10.20& 56.9& 0.900 & 0.845 \\
        \textbf{1.0}   & \textbf{91.7} & \textbf{7.24} & \textbf{45.8} & \textbf{0.954} & \textbf{0.913} \\
        1.1  & 77.8 & 9.73 & 52.0 & 0.949 & 0.895 \\
        1.5 & 54.2 & 64.55 & 87.8 & 0.178 & 0.497 \\
        \toprule
    \end{tabular}
    
    \label{table:temp}
\end{table}

\subsection{Ablation Study on Number of Codebooks}
As discussed earlier, the proposed method is inherently constrained by the number of codebooks employed for modeling. As shown in Table \ref{table:result}, restoration quality is strongly correlated with the number of codebooks, where using fewer codebooks results in increasingly severe reconstruction distortion. Theoretically, optimal performance would be achieved by fully leveraging Encodec’s complete 32-order codebook sequence. However, due to the use of independent token embedding layers for different orders in the NAR model, predicting higher-order codebooks requires jointly integrating information from all lower-order codebooks. This design substantially increases training complexity and computational cost, leading to significantly slower convergence in practice. After carefully balancing training efficiency, computational overhead, and reconstruction quality, we finally select the top 8 codebooks for modeling in our main experiments. This choice represents a practical compromise that preserves most of the perceptual and linguistic benefits while maintaining a feasible training cost.

\begin{figure}[!htb]
    \centering
    \includegraphics[width=1\linewidth]{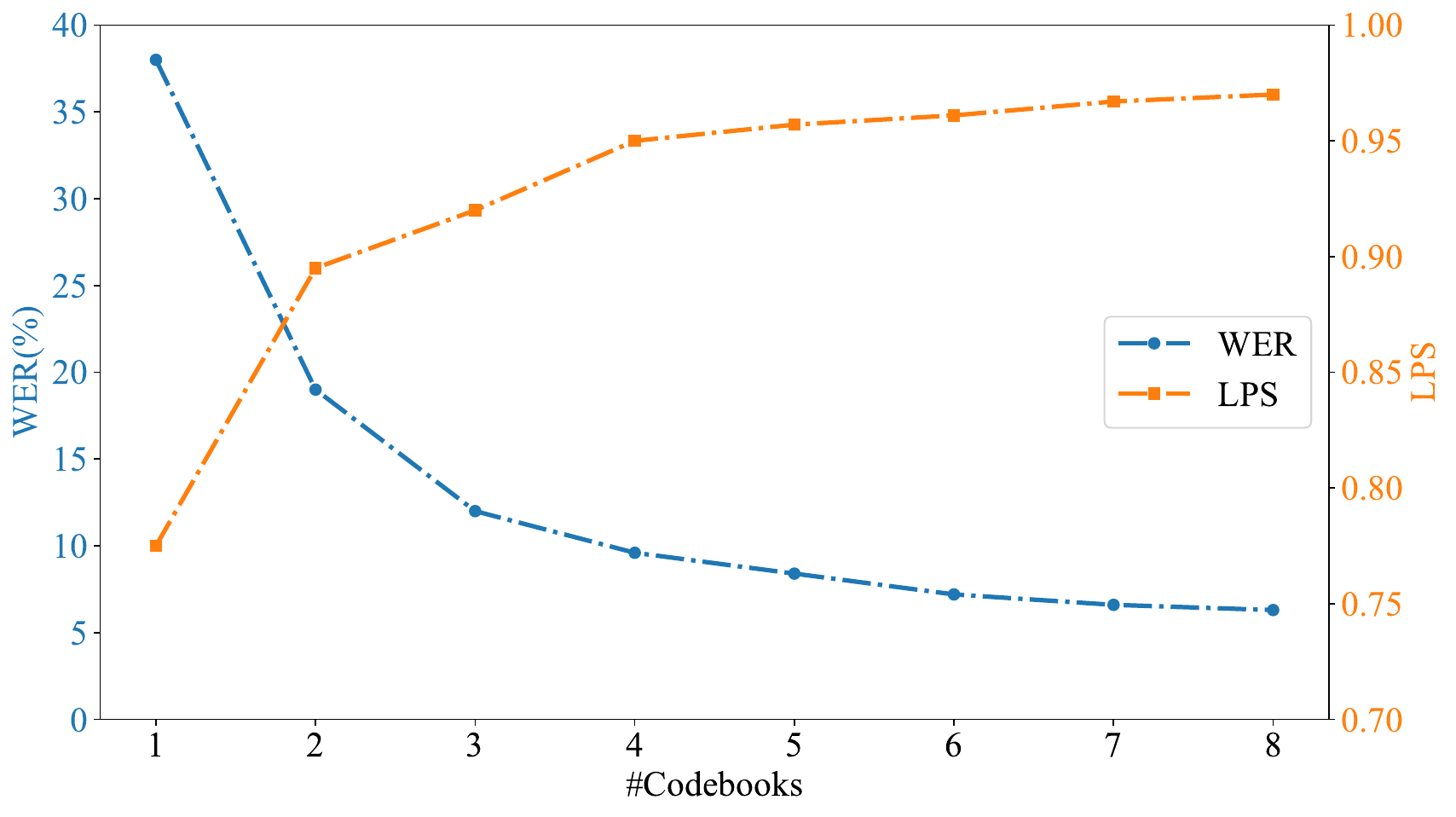}
    \caption{Performance variation of WER and LPS with the number of codebooks. The codebook exhibits distinct residual characteristics, and its performance changes nonlinearly; when the number of codebooks exceeds 6, the improvement in performance gradually slows down.}
    \label{fig:abla}
\end{figure}

To quantitatively analyze this relationship, we evaluate multiple metrics using the first 1 to 7 codebooks and illustrate the corresponding performance trends in Fig. \ref{fig:abla}. As the number of codebooks increases, model performance exhibits a clear and consistent positive correlation across all evaluated metrics. These results indicate that incorporating additional codebooks can further enhance the overall performance of our method.

\subsection{Performance on Scenarios with Dynamic Numbers of Talkers}
We further investigate the performance of the SLM-SS framework under scenarios involving different numbers of talkers. Specifically, we train and evaluate the model on Libri2Mix, Libri3Mix, and their joint training set to assess robustness to varying numbers of speakers. Due to computational constraints in this experiment, we replace the Whisper-medium configuration with Whisper-base, while keeping all other training hyperparameters unchanged. This experimental setting enables a controlled comparison across different talker scenarios without introducing additional confounding factors. The corresponding results are reported in Table \ref{table:spknum}.

\begin{table}[!htp]
\centering
\caption{Performance with different numbers of talkers.}
\label{table:spknum}
\begin{tabular}{llccccc}
\toprule
\textbf{Train} & \textbf{Test} & \textbf{Spk sim} & \textbf{WER} & \textbf{TER} & \textbf{LPS}  &  \textbf{SBS}  \\
\midrule

\multirow{1}{*}{\textbf{2Mix}}
& 2Mix & 90.2 & 8.84 & 51.1 & 0.936 & 0.900 \\
\midrule

\multirow{1}{*}{\textbf{3Mix}}
& 3Mix & 87.3 & 12.12 & 63.7 & 0.895 & 0.870\\
\midrule

\multirow{2}{*}{\textbf{Joint}}
& 2Mix & 90.2 & 8.80 & 50.7 & 0.939 & 0.903 \\
\cmidrule(lr){2-7}
& 3Mix & 88.8 & 9.94 & 61.3 & 0.915 & 0.875 \\
\bottomrule
\end{tabular}
\end{table}

Table \ref{table:spknum} reveals a clear talker-count effect: compared with 2Mix, performance degrades consistently in the 3Mix-only setting across all objective metrics, confirming that three-talker separation is substantially more challenging. This trend is expected because higher-talker mixtures produce longer SOT sequences, which increase autoregressive modeling difficulty. More importantly, joint training on 2Mix and 3Mix improves robustness in the harder 3Mix condition, with a clear WER reduction ($12.1\%$$\rightarrow$ $9.9\%$) and consistent gains on the remaining metrics, while preserving performance in the easier 2Mix condition. Overall, mixed-talker training provides a better robustness-accuracy trade-off than training on a single talker-count setting.

\subsection{Robustness Under Increasing Number of Talker}

To evaluate the scalability of the SLM-SS to more challenging mixtures, we extend the LibriMix data simulation pipeline to construct a 4-talker setting. With all hyperparameters kept unchanged, we generate a Libri4Mix dataset (with a small-scale train set and a test set) and finetune the model initialized from AR weights jointly trained on 2 and 3 talker scenarios.

\begin{figure}[!htb]
\centering
\includegraphics[width=1\linewidth]{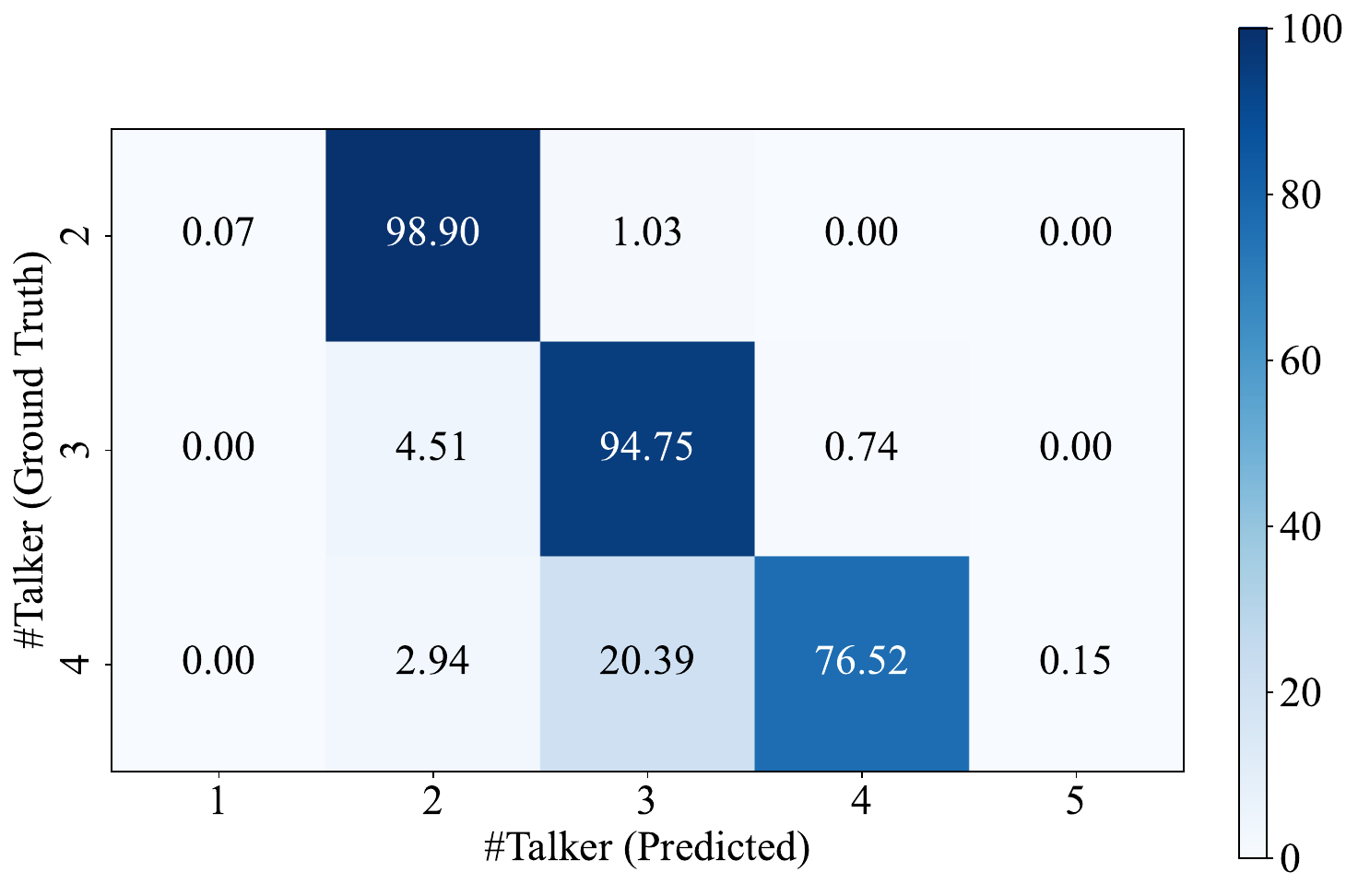}
\caption{Confusion matrix for talker number prediction accuracy across different scenarios.}
\label{fig:spkcnt}
\end{figure}

The model converges within three epochs under the limited 4-talker data, after which the validation loss increases, indicating overfitting. This suggests that the framework can be effectively adapted to higher talker-count scenarios via lightweight finetuning, even under low-resource conditions. We further assess talker number prediction across different scenarios by applying the fine-tuned 4-talker model to all settings. The resulting confusion matrices are shown in Fig.~\ref{fig:spkcnt}.

The performance on 2- and 3-talker scenarios remains largely unaffected after finetuning, with only a small number of prediction errors, due to early stopping preventing severe overfitting. In contrast, performance in the 4-talker scenario is lower due to increased task difficulty and limited data, with occasional underestimation of talker counts. Nevertheless, the model still demonstrates strong overall separation capability, highlighting the robustness of the proposed framework in more challenging multi-talker settings.

Furthermore, we adopt the per-talker NAR approach for high-order sequence decoding to evaluate the quality of the final audio output. The NAR model, jointly trained on 2 and 3 talker scenarios, is finetuned using the same strategy as the AR component and evaluated across different talker settings. Additionally, to investigate whether the per-talker NAR model possesses the generalization capability to directly apply to four-talker scenarios, we also conducted direct tests using the unfinetuned weights.

\begin{table}[!htp]
\centering
\caption{Performance Comparison Across Different Numbers of Talkers After Extension}
\label{table:spk4}
\resizebox{\columnwidth}{!}{%
\begin{tabular}{@{}ccccccc@{}}
\toprule
\textbf{NAR} & \textbf{\#Talker} & \textbf{Spk Sim.$\uparrow$ } & \textbf{WER}$\downarrow$ & \textbf{TER}$\downarrow$ & \textbf{LPS}$\uparrow$  &  \textbf{SBS}$\uparrow$  \\
\midrule

\multirow{3}{*}{\textbf{Unfinetuned}}
& \textbf{2Mix} & 90.1 & 8.83 & 50.5 & 0.937 & 0.903 \\
\cmidrule(lr){2-7}
& \textbf{3Mix} & 88.7 & 9.93 & 61.3 & 0.912 & 0.871 \\
\cmidrule(lr){2-7}
& \textbf{4Mix} & 79.5 & 16.19 & 70.1 & 0.833 & 0.794 \\

\midrule

\multirow{3}{*}{\textbf{Finetuned}}
& \textbf{2Mix} & 90.1 & 8.83 & 50.6 & 0.938 & 0.903 \\
\cmidrule(lr){2-7}
& \textbf{3Mix} & 88.7 & 9.92 & 61.4 & 0.911 & 0.872 \\
\cmidrule(lr){2-7}
& \textbf{4Mix} & 84.6 & 11.10 & 66.7 & 0.889 & 0.852 \\

\bottomrule
\end{tabular}%
}
\end{table}

As shown in Table \ref{table:spk4}, when the NAR model is not finetuned on the 4-talker data, performance in the 2 and 3 talker scenarios is largely governed by the AR component finetuned for the 4-talker setting, with no significant performance changes observed. However, in the 4-talker scenario, the original per-talker NAR encoder lacks the capacity to model more complex mixtures, leading to insufficient extraction of higher-order codebooks and consequent performance degradation, but it still offers a certain level of performance.

In contrast, after finetuning the per-talker NAR model, performance in the 2 and 3 talker scenarios remains stable, while performance in the 4-talker scenario improves substantially. These results indicate a strong correlation between the number of talkers (i.e., inference difficulty) and model performance, demonstrating that the SLM-SS framework effectively supports variable numbers of talkers and maintains consistent performance across different speaker configurations.

\subsection{Results on AR-Only Model}
To study the effect of the neural codec choice, we replace Encodec with WavTokenizer while keeping the AR backbone and training recipe unchanged. Concretely, we only adjust the vocabulary size from 1024 to 4096 to match WavTokenizer; all other hyperparameters remain identical to ensure a fair comparison. The AR model is trained separately on 2- and 3-talker mixtures.

As summarized in the last row of Table~\ref{table:result}, switching from Encodec to WavTokenizer yields comparable separation quality (e.g., SBS) while slightly degrading WER. We attribute this trend to the codec objective: WavTokenizer is more aggressively optimized at the signal level, which benefits perceptual quality and stability, whereas it is not explicitly tailored for ASR-related objectives.

Beyond the metric comparison, WavTokenizer also changes the system design trade-off. In our framework, it removes the need for the additional per-talker NAR module for high-order codebook decoding, which simplifies the training/inference pipeline and reduces the overall compute cost.

\subsection{Decoding Strategies with AR-Only Model}
We further analyze the impact of autoregressive decoding under WavTokenizer in Table~\ref{table:wavtokenizer}. We consider greedy decoding, beam search ($beamsize=20$), and a lightweight stochastic variant, Top-30 Random Sampling (Top-30 RS): at each step, we first select the 30 tokens with the highest logits and then uniformly sample one token from this candidate set.

\begin{table}[!htp]
\centering
\caption{Performance using WavTokenizer under different talker scenarios and decoding strategies.}
\label{table:wavtokenizer}
\begin{tabular}{@{}llcccc@{}}
\toprule
\textbf{Dataset} & \textbf{Decode method} & \textbf{Spk Sim.}$\uparrow$ & \textbf{WER}$\downarrow$ & \textbf{LPS} $\uparrow$ &  \textbf{SBS}$\uparrow$  \\
\midrule

\multirow{3}{*}{\textbf{2Mix}}
& Greedy Search & 91.9 & 9.66 & 0.904 & 0.915 \\
\cmidrule(lr){2-6}
& Top-30 RS & 92.0 & 9.08 & 0.917 & 0.919  \\
\cmidrule(lr){2-6}
& Beam Search   & 92.3 & 8.97 & 0.923 & 0.921  \\
\midrule

\multirow{3}{*}{\textbf{3Mix}}
& Greedy Search & 87.2 & 11.41 & 0.906 & 0.910 \\
\cmidrule(lr){2-6}
& Top-30 RS & 87.8 & 11.03 & 0.908 & 0.911\\
\cmidrule(lr){2-6}
& Beam Search & 88.8 & 10.18 & 0.915 & 0.913  \\
\bottomrule
\end{tabular}%
\end{table}

Compared with Top-1 greedy decoding, Top-30 RS provides a more robust decoding behavior (consistent gains on both 2Mix and 3Mix), likely because the larger and more redundant WavTokenizer vocabulary makes the model less sensitive to occasional over-confident token choices. Beam search still delivers the best overall performance, indicating the expected trade-off between decoding efficiency and optimality.

Overall, the codec choice affects both the achievable performance and the computational profile of SLM-SS. Exploring more advanced neural codecs (and their matching decoding strategies) remains a promising direction for future work.

\subsection{Ablation Study on Number of Model Parameters}
\label{sec:params}
Since the codebook sequence length increases linearly with the number of talkers, SLM require sufficient capacity for autoregressive decoding. Insufficient capacity may cause cumulative errors over long sequences, whereas excessive capacity substantially increases computational overhead.

To investigate the trade-off between model size and performance, we evaluated two pre-trained encoder sizes and four scratch-built decoder sizes for SLM-SS with the WavTokenizer-based approach. We report performance, parameter count, computational cost (GMACs) at 16 kHz, peak memory allocation, and Real-time Factor (RTF), and compare them with two advanced discriminative models. GMACs are computed per second of 16-kHz input audio. RTF
and peak allocated GPU memory are measured on an RTX 4090 using FP16 inference with batch size 1. The reported RTF includes autoregressive beam-search decoding with $beamsize=20$.

As shown in Table \ref{table:model_scaling}, the encoder extracts latent
representations from multi-talker audio and guides the
autoregressive decoder through cross-attention. Insufficient encoder capacity therefore limits decoding performance and cannot be fully compensated for by a larger decoder. With a sufficiently powerful encoder, a moderately sized decoder achieves strong performance, while an overly small decoder may struggle with long sequences. Further increasing decoder size yields diminishing gains with substantially higher computational overhead.

Despite having fewer parameters, the computational costs of discriminative models remain high and are comparable to those of our medium-sized SLM model.
They directly predict masks or spectrum mappings without autoregressive decoding, achieving much lower RTF than all SLM configurations. This holds even when some discriminative models requires higher overall computational overhead than SLM models. 
Due to the non-causal encoder and full-context cross-attention, the current SLM-SS implementation is intended for offline inference.

\begin{table*}[!t]
\centering
\caption{Model scaling and computational efficiency comparison.}
\label{table:model_scaling}

\small
\setlength{\tabcolsep}{4pt}
\renewcommand{\arraystretch}{1.12}

\begin{tabular*}{\textwidth}{@{\extracolsep{\fill}}lcccccccc@{}
}
\toprule

\textbf{Configuration}
& \shortstack{\textbf{\#Params}}
& \shortstack{\textbf{GMACs}} & \textbf{Mem} & \textbf{RTF} & \textbf{WER}$\downarrow$ & \textbf{PESQ}$\uparrow$ & \textbf{UTMOS}$\uparrow$ & \textbf{DNSMOS}$\uparrow$
\\
\midrule
\multicolumn{8}{l}{
    \textbf{WavLM Base+ Encoder (94M)}
}
\\
\hspace{1.5em}+ 23M Decoder (b-t)
& 117M & 37.14 & 0.72GiB & 0.29 & 72.87 & 1.22 & 2.21 & 3.07 \\
\hspace{1.5em}+ 48M Decoder (b-s)
& 142M & 69.52 & 1.44GiB & 0.52 & 63.88 & 1.34 & 2.58 & 3.16 \\
\hspace{1.5em}+ 121M Decoder (b-m)
& 215M & 161.01 & 2.18GiB & 0.75 & 36.91 & 1.65 & 3.01 & 3.29 \\
\hspace{1.5em}+ 212M Decoder (b-l)
& 306M & 348.73 & 2.90GiB & 0.91 & 25.22 & 1.91 & 3.32 & 3.40 \\
\cmidrule(lr){1-9}
\multicolumn{8}{l}{
    \textbf{WavLM Large Encoder (315M)}
}
\\
\hspace{1.5em}+ 23M Decoder (l-t)
& 338M & 45.82 & 1.47GiB & 0.37 & 46.29 & 1.52 & 2.40 & 3.11 \\
\hspace{1.5em}+ 48M Decoder (l-s) & 363M & 84.56 & 2.19GiB & 0.64 & 9.43 & 2.07 & \textbf{3.97} & 3.61\\
\hspace{1.5em}+ 121M Decoder (l-m)
& 436M & 177.83 & 2.93GiB & 0.93 & 10.14 & 2.06 & 3.94 & \textbf{3.66} \\
\hspace{1.5em}+ 212M Decoder (l-l, \textbf{ours})
& 527M & 366.39 & 3.66GiB & 1.35 & 8.97 & 2.06 & 3.95 & 3.65 \\
\midrule
\textbf{TF-LocoFormer}
& 15M & 250.20 & 1.19GiB & 0.08 & \textbf{5.88} & \textbf{3.51} & 3.79 & 3.56 \\
\textbf{MossFormer2}
& 53M & 105.06 & 1.44GiB & 0.07 & 8.78 & 2.95 & 3.27 & 3.24 \\
\bottomrule
\end{tabular*}
\end{table*}

\subsection{Performance on Out-of-domain Data}

To further investigate the generalization ability of the SLM-SS model, we selected two out-of-domain datasets: the Libri2Mix noisy testset, which incorporates WHAM!~\cite{wham} noise; and the AISHELL1-2Mix Chinese noise-free dataset, generated based on AISHELL1~\cite{aishell1} using a simulation pipeline similar to the Librimix. We selected models from the \textit{l-l} and \textit{l-s} configurations in Table \ref{table:model_scaling}, both of which were trained on the clean LibriMix dataset, to evaluate their cross-lingual and noise-robust capabilities. For comparison, we also tested TF-LocoFormer and MossFormer2, which were trained on the same LibriMix dataset.

\begin{table}[!t]
\centering
\caption{Out-of-domain evaluation on Libri2Mix-noisy and AISHELL1-2Mix.}
\label{table:zero-shot}

\begin{tabular*}{\linewidth}{
    @{\extracolsep{\fill}}
    lcccc
    @{}
}
\toprule
\textbf{Model}
& \textbf{WER/CER}$\downarrow$
& \textbf{PESQ}$\uparrow$
& \textbf{UTMOS}$\uparrow$
& \textbf{DNSMOS}$\uparrow$\\
\midrule

\multicolumn{5}{c}{\textit{(a) Libri2Mix-noisy}}\\
\midrule

SLM-SS-WT (l-s)
& 30.46
& 1.60
& 2.96
& 3.20\\

SLM-SS-WT (l-l)
& \textbf{25.54}
& 1.61
& \textbf{2.98}
& \textbf{3.22}\\

TF-LocoFormer
& 25.96
& \textbf{1.64}
& 2.11
& 2.56\\

MossFormer2
& 33.89
& 1.39
& 1.54
& 2.35\\

\midrule
\multicolumn{5}{c}{\textit{(b) AISHELL1-2Mix}}\\
\midrule

SLM-SS-WT (l-s)
& 38.35$^{*}$
& 1.56
& 2.65
& \textbf{3.48}\\

SLM-SS-WT (l-l)
& 33.70$^{*}$
& 1.55
& \textbf{2.66}
& \textbf{3.48}\\

TF-LocoFormer
& \textbf{32.58}$^{*}$
& \textbf{2.25}
& 1.82
& 2.91\\

MossFormer2
& 47.36$^{*}$
& 1.77
& 1.46
& 2.90\\

\bottomrule
\end{tabular*}

{\footnotesize
\raggedright
WER is reported for Libri2Mix-noisy, while CER is reported for AISHELL1-2Mix.
\par}
\end{table}

As shown in Table \ref{table:zero-shot}, SLM-SS achieves the best DNSMOS and UTMOS on both out-of-domain datasets. In terms of WER, the \textit{l-l} configuration outperforms the discriminative models on LibriMix noisy, while performing slightly worse than TF-LocoFormer on the Chinese dataset, demonstrating the cross-language generalization and noise robustness of the SLM models.


\section{Discussion}
\label{sec:discussion}

\subsection{Neural codecs are the dominant performance bottleneck.}
Our experiments indicate that, in Encodec-based AR--NAR separation, the effective performance ceiling is the codec reconstruction from the modeled codebooks (GT-Encodec8), rather than the original waveform. Consequently, codec choice and how its representation is exploited largely determine the attainable separation quality. Under residual vector quantization, adding higher-order codebooks helps (Fig.~\ref{fig:abla}) but substantially complicates optimization, since each higher-order prediction must condition on an expanding set of lower-order sequences---highlighting a trade-off between information completeness and trainability/efficiency.

WavTokenizer further shows that codec design is a first-order algorithmic decision: a single-order codec can remove the NAR refinement stage, simplifying training/inference and improving subjective quality. Yet its objective (often waveform/perception oriented) may mismatch ASR-centric intelligibility metrics, yielding different WER--MOS trade-offs. Overall, progress may depend more on co-designing the codec and token-prediction objective for separation and intelligibility than on scaling the SLM backbone alone.

\subsection{Why use continuous mixture inputs with discrete token outputs?}
We adopt a continuous encoder with a discrete token decoder because current neural codecs are predominantly trained on single-speaker speech, and thus are not guaranteed to produce a well-behaved discrete representation when directly encoding multi-talker mixtures. If we were to feed mixture waveforms into a codec and then run a purely decoder-only token model, the discretization step itself could become a major failure point: mixture interference may be absorbed into the codec tokens in an entangled manner, making it difficult to recover speaker-consistent sequences.

By keeping the mixture processing in a continuous encoder, the attention-based encoder-decoder model can learn separation-relevant acoustic representations directly from the mixture, while using discrete tokens only on the output side as an intelligibility-oriented generative target. This hybrid design also keeps the framework compatible with future codecs: if a high-quality multi-talker codec (or a mixture-invariant, speaker-disentangling tokenizer) becomes available, our framework could be naturally simplified into a decoder-only architecture that predicts tokens conditioned on discretized mixture inputs, further reducing system complexity and latency.

\subsection{Limitations and future work}
\noindent \textbf{Codec--model co-design.} The strongest path forward is likely to adopt codecs with fewer codebooks but higher information density, or to learn task-specific tokenizers that better preserve intelligibility-relevant cues under aggressive compression. Jointly training (or at least jointly selecting) codec representations and separation objectives may reduce the current gap between perceptual quality and downstream intelligibility.

\noindent\textbf{Scaling to more talkers.} Higher talker counts lengthen the SOT sequence and increase uncertainty in speaker-change/boundary decisions, making AR decoding more sensitive to early token errors (e.g., boundary shifts, speaker swaps, or incorrect talker counts). While finetuning can extend the framework (e.g., to 4 talkers), performance is quickly limited by data scarcity at higher talker counts; addressing this likely requires more diverse high-overlap mixtures and more robust decoding constraints or auxiliary objectives (e.g., on \texttt{<SC>} placement) to stabilize boundary prediction.

\noindent\textbf{Real-world robustness.} LibriMix is simulated and largely clean; it does not fully cover realistic reverberation, background noise, channel mismatch, and spontaneous speech. Extending SLM-SS to such conditions likely requires either domain-augmented training or explicit modeling of acoustics/noise in the token space.

\section{Conclusion}
\label{sec:conclusion}
We presented SLM-SS, a token-based generative speech separation framework that leverages speech language models. Clean targets are discretized into multi-codebook sequences with neural codecs and serialized with SOT, turning multi-talker separation into sequence generation. An encoder--decoder AR model predicts the zero-order codebook to establish speaker ordering and coarse linguistic content, and a lightweight per-talker NAR module then refines higher-order residual codebooks for efficient reconstruction.

Experiments show that per-talker refinement improves recognition accuracy, while the single-codebook WavTokenizer variant simplifies generation and improves listener-rated quality. Its non-intrusive quality advantages over the tested baselines persist on noisy and Chinese mixtures. Together, these results demonstrate SLM-SS's ability to recover intelligible, perceptually high-quality speech from multi-talker mixtures through codec-token generation. Future work will focus on codec--model co-design, faster decoding, and robustness to more challenging multi-talker conditions.

\bibliographystyle{named}
\bibliography{ref}

@inproceedings{tangTSELMTargetSpeaker2026,
  title = {{{TSELM}}: {{Target Speaker Extraction Using Discrete Tokens}} and~{{Language Models}}},
  shorttitle = {{{TSELM}}},
  booktitle = {Man-{{Machine Speech Communication}}},
  author = {Tang, Beilong and Zeng, Bang and Li, Ming},
  editor = {Jia, Jia and Wu, Zhiyong and Gao, Lijian and Huang, Gongping and Li, Ya},
  year = 2026,
  pages = {459--469},
  publisher = {Springer Nature},
  address = {Singapore},
  doi = {10.1007/978-981-95-5382-2_35},
  isbn = {9789819553822},
  langid = {english},
}

@article{wangComplexSpectralMapping2020,
  title = {Complex {{Spectral Mapping}} for {{Single-}} and {{Multi-Channel Speech Enhancement}} and {{Robust ASR}}},
  author = {Wang, Zhong-Qiu and Wang, Peidong and Wang, DeLiang},
  year = 2020,
  journal = {IEEE/ACM Transactions on Audio, Speech, and Language Processing},
  volume = {28},
  pages = {1778--1787},
  issn = {2329-9304},
  doi = {10.1109/TASLP.2020.2998279}
}

@inproceedings{luoDualPathRNNEfficient2020,
  title = {Dual-{{Path RNN}}: {{Efficient Long Sequence Modeling}} for {{Time-Domain Single-Channel Speech Separation}}},
  shorttitle = {Dual-{{Path RNN}}},
  booktitle = {{{ICASSP}} 2020 - 2020 {{IEEE International Conference}} on {{Acoustics}}, {{Speech}} and {{Signal Processing}} ({{ICASSP}})},
  author = {Luo, Y. and Chen, Z. and Yoshioka, T.},
  year = 2020,
  month = may,
  pages = {46--50},
  address = {Barcelona},
  issn = {2379-190X},
  doi = {10.1109/ICASSP40776.2020.9054266}
}

@inproceedings{liSkimSkippingMemory2022,
  title = {Skim: {{Skipping Memory Lstm}} for {{Low-Latency Real-Time Continuous Speech Separation}}},
  shorttitle = {Skim},
  booktitle = {{{ICASSP}} 2022 - 2022 {{IEEE International Conference}} on {{Acoustics}}, {{Speech}} and {{Signal Processing}} ({{ICASSP}})},
  author = {Li, Chenda and Yang, Lei and Wang, Weiqin and Qian, Yanmin},
  year = 2022,
  month = may,
  pages = {681--685},
  address = {Singapore},
  issn = {2379-190X},
  doi = {10.1109/ICASSP43922.2022.9746372}
}

@inproceedings{chenDualPathTransformerNetwork2020,
  title = {Dual-{{Path Transformer Network}}: {{Direct Context-Aware Modeling}} for {{End-to-End Monaural Speech Separation}}},
  shorttitle = {Dual-{{Path Transformer Network}}},
  booktitle = {Interspeech 2020},
  author = {Chen, Jingjing and Mao, Qirong and Liu, Dong},
  year = 2020,
  month = oct,
  pages = {2642--2646},
  publisher = {ISCA},
  doi = {10.21437/Interspeech.2020-2205},
  urldate = {2021-04-16},
  langid = {english}
}

@inproceedings{lanzendorferHighFidelitySpeechEnhancement2026,
  title = {High-{{Fidelity Speech Enhancement Via Discrete Audio Tokens}}},
  booktitle = {{{ICASSP}} 2026 - 2026 {{IEEE International Conference}} on {{Acoustics}}, {{Speech}} and {{Signal Processing}} ({{ICASSP}})},
  author = {Lanzend{\"o}rfer, Luca A. and Berdoz, Fr{\'e}d{\'e}ric and Asonitis, Antonis and Wattenhofer, Roger},
  year = 2026,
  month = may,
  pages = {17397--17401},
  issn = {2379-190X},
  doi = {10.1109/ICASSP55912.2026.11460900},
  urldate = {2026-05-12}
}

@article{zhouUniVoiceUnifiedFramework2026,
  title = {{{UniVoice}}: A Unified Framework for Text-to-Speech, Singing Voice Synthesis, and Opera Singing Synthesis},
  author = {Zhou, Yue and Bai, Peng and Shi, Xiaodong},
  year = 2026,
  month = feb,
  journal = {Applied Intelligence},
  volume = {56},
  number = {4},
  pages = {104},
  issn = {1573-7497},
  doi = {10.1007/s10489-026-07137-w},
}

@misc{shiQwen3ASRTechnicalReport2026,
  title = {Qwen3-{{ASR Technical Report}}},
  author = {Shi, Xian and Wang, Xiong and Guo, Zhifang and Wang, Yongqi and Zhang, Pei and Zhang, Xinyu and Guo, Zishan and Hao, Hongkun and Xi, Yu and Yang, Baosong and Xu, Jin and Zhou, Jingren and Lin, Junyang},
  year = 2026,
  month = jan,
  number = {arXiv:2601.21337},
  eprint = {2601.21337},
  primaryclass = {cs.CL},
  publisher = {arXiv},
  doi = {10.48550/arXiv.2601.21337},
  urldate = {2026-05-14},
  archiveprefix = {arXiv},
}

@article{vasukiReviewVectorQuantization2006,
  title = {A Review of Vector Quantization Techniques},
  author = {Vasuki, A. and Vanathi, P.T.},
  year = 2006,
  month = jul,
  journal = {IEEE Potentials},
  volume = {25},
  number = {4},
  pages = {39--47},
  issn = {1558-1772},
  doi = {10.1109/MP.2006.1664069},
  urldate = {2026-05-14}
}

@article{grayVectorQuantization1984,
  title = {Vector Quantization},
  author = {Gray, R.},
  year = 1984,
  month = apr,
  journal = {IEEE ASSP Magazine},
  volume = {1},
  number = {2},
  pages = {4--29},
  issn = {1558-1284},
  doi = {10.1109/MASSP.1984.1162229},
  urldate = {2026-05-14}
}

@article{wangUSEUnifiedModel2026,
  title = {{{USE}}: {{A Unified Model}} for {{Universal Sound Separation}} and {{Extraction}}},
  shorttitle = {{{USE}}},
  author = {Wang, Hongyu and Li, Chenda and Zhou, Xin and Wang, Shuai and Qian, Yanmin},
  year = 2026,
  month = mar,
  journal = {Proceedings of the AAAI Conference on Artificial Intelligence},
  volume = {40},
  number = {39},
  pages = {33476--33484},
  issn = {2374-3468},
  doi = {10.1609/aaai.v40i39.40635},
  urldate = {2026-04-14},
  copyright = {Copyright (c) 2026 Association for the Advancement of Artificial Intelligence},
  langid = {english},
}

@inproceedings{liSLMSSSpeechLanguage2026,
  title = {{{SLM-SS}}: {{Speech Language Model}} for {{Generative Speech Separation}}},
  shorttitle = {{{SLM-SS}}},
  booktitle = {{{ICASSP}} 2026 - 2026 {{IEEE International Conference}} on {{Acoustics}}, {{Speech}} and {{Signal Processing}} ({{ICASSP}})},
  author = {Li, Tianhua and Li, Chenda and Wang, Wei and Zhou, Xin and Chen, Xihui and Gao, Jianqing and Qian, Yanmin},
  year = 2026,
  month = may,
  pages = {15432--15436},
  issn = {2379-190X},
  doi = {10.1109/ICASSP55912.2026.11463173},
  urldate = {2026-04-27}
}

@inproceedings{wangMeanSEEfficientGenerative2026,
  title = {{{MeanSE}}: {{Efficient Generative Speech Enhancement}} with {{Mean Flows}}},
  shorttitle = {{{MeanSE}}},
  booktitle = {{{ICASSP}} 2026 - 2026 {{IEEE International Conference}} on {{Acoustics}}, {{Speech}} and {{Signal Processing}} ({{ICASSP}})},
  author = {Wang, Jiahe and Wang, Hongyu and Wang, Wei and Yang, Lei and Li, Chenda and Zhang, Wangyou and Tan, Lufen and Qian, Yanmin},
  year = 2026,
  month = may,
  pages = {16827--16831},
  issn = {2379-190X},
  doi = {10.1109/ICASSP55912.2026.11464536},
  urldate = {2026-04-27},
}

@inproceedings{scheiblerSourceSeparationFlow2025a,
  title = {Source {{Separation}} by {{Flow Matching}}},
  booktitle = {2025 {{IEEE Workshop}} on {{Applications}} of {{Signal Processing}} to {{Audio}} and {{Acoustics}} ({{WASPAA}})},
  author = {Scheibler, Robin and Hershey, John R. and Doucet, Arnaud and Li, Henry},
  year = 2025,
  month = oct,
  pages = {1--5},
  issn = {1947-1629},
  doi = {10.1109/WASPAA66052.2025.11230963},
  urldate = {2026-05-12}
}

@ARTICLE{ss-survey,
  author={Wang, DeLiang and Chen, Jitong},
  journal={IEEE/ACM Transactions on Audio, Speech, and Language Processing}, 
  title={Supervised Speech Separation Based on Deep Learning: An Overview}, 
  year={2018},
  volume={26},
  number={10},
  pages={1702-1726},
  doi={10.1109/TASLP.2018.2842159}}

@INPROCEEDINGS{sisdr,
  author={Roux, Jonathan Le and Wisdom, Scott and Erdogan, Hakan and Hershey, John R.},
  booktitle={ICASSP 2019 - 2019 IEEE International Conference on Acoustics, Speech and Signal Processing (ICASSP)}, 
  title={SDR – Half-baked or Well Done?}, 
  year={2019},
  volume={},
  number={},
  pages={626-630},
  doi={10.1109/ICASSP.2019.8683855}}

@inproceedings{distortion,
  title     = {{How bad are artifacts?: Analyzing the impact of speech enhancement errors on ASR}},
  author    = {Kazuma Iwamoto and Tsubasa Ochiai and Marc Delcroix and Rintaro Ikeshita and Hiroshi Sato and Shoko Araki and Shigeru Katagiri},
  year      = {2022},
  booktitle = {{Interspeech 2022}},
  pages     = {5418--5422},
  doi       = {10.21437/Interspeech.2022-318},
  issn      = {2958-1796},
}

@ARTICLE{deepcasa,
  author={Liu, Yuzhou and Wang, DeLiang},
  journal={IEEE/ACM Transactions on Audio, Speech, and Language Processing}, 
  title={Divide and Conquer: A Deep CASA Approach to Talker-Independent Monaural Speaker Separation}, 
  year={2019},
  volume={27},
  number={12},
  pages={2092-2102},
  doi={10.1109/TASLP.2019.2941148}}

@INPROCEEDINGS{deepclustering,
  author={Hershey, John R. and Chen, Zhuo and Le Roux, Jonathan and Watanabe, Shinji},
  booktitle={2016 IEEE International Conference on Acoustics, Speech and Signal Processing (ICASSP)}, 
  title={Deep clustering: Discriminative embeddings for segmentation and separation}, 
  year={2016},
  volume={},
  number={},
  pages={31-35},
  doi={10.1109/ICASSP.2016.7471631}}

@inproceedings{hallucination,
 author = {Aithal, Sumukh K and Maini, Pratyush and Lipton, Zachary C. and Kolter, J. Zico},
 booktitle = {Advances in Neural Information Processing Systems},
 doi = {10.52202/079017-4278},
 editor = {A. Globerson and L. Mackey and D. Belgrave and A. Fan and U. Paquet and J. Tomczak and C. Zhang},
 pages = {134614--134644},
 publisher = {Curran Associates, Inc.},
 title = {Understanding Hallucinations in Diffusion Models through Mode Interpolation},
 url = {https://proceedings.neurips.cc/paper_files/paper/2024/file/f29369d192b13184b65c6d2515474d78-Paper-Conference.pdf},
 volume = {37},
 year = {2024}
}

@inproceedings{tasnet-luo2018,
author = {Luo, Yi and Mesgarani, Nima},
title = {TaSNet: Time-Domain Audio Separation Network for Real-Time, Single-Channel Speech Separation},
year = {2018},
publisher = {IEEE Press},
doi = {10.1109/ICASSP.2018.8462116},
booktitle = {2018 IEEE International Conference on Acoustics, Speech and Signal Processing (ICASSP)},
pages = {696–700},
numpages = {5},
location = {Calgary, AB, Canada}
}

@ARTICLE{conv-tasnet,
  author={Luo, Yi},
title={Conv-TasNet: Surpassing Ideal Time–Frequency Magnitude Masking for Speech Separation}, 
  journal={IEEE/ACM Transactions on Audio, Speech, and Language Processing}, 
  year={2019},
  volume={27},
  number={8},
  pages={1256-1266},
  doi={10.1109/TASLP.2019.2915167}}

@ARTICLE{bsrnn-luo2023,
  author={Luo, Yi and Yu, Jianwei},
  journal={IEEE/ACM Transactions on Audio, Speech, and Language Processing}, 
  title={Music Source Separation With Band-Split RNN}, 
  year={2023},
  volume={31},
  number={},
  pages={1893-1901},
  doi={10.1109/TASLP.2023.3271145}}

@INPROCEEDINGS{sepformer-cem2021,
  author={Subakan, Cem and Ravanelli, Mirco and Cornell, Samuele and Bronzi, Mirko and Zhong, Jianyuan},
  booktitle={ICASSP 2021 - 2021 IEEE International Conference on Acoustics, Speech and Signal Processing (ICASSP)}, 
  title={Attention Is All You Need In Speech Separation}, 
  year={2021},
  volume={},
  number={},
  pages={21-25},
  doi={10.1109/ICASSP39728.2021.9413901}}

@INPROCEEDINGS{tuss,
  author={Saijo, Kohei and Ebbers, Janek and Germain, François G. and Wichern, Gordon and Le Roux, Jonathan},
  booktitle={ICASSP 2025 - 2025 IEEE International Conference on Acoustics, Speech and Signal Processing (ICASSP)}, 
  title={Task-Aware Unified Source Separation}, 
  year={2025},
  volume={},
  number={},
  pages={1-5},
  doi={10.1109/ICASSP49660.2025.10887819}}

@INPROCEEDINGS{flexio,
  author={Masuyama, Yoshiki and Saijo, Kohei and Paissan, Francesco and Han, Jiangyu and Delcroix, Marc and Aihara, Ryo and Germain, François G. and Wichern, Gordon and Le Roux, Jonathan},
  booktitle={ICASSP 2026 - 2026 IEEE International Conference on Acoustics, Speech and Signal Processing (ICASSP)}, 
  title={Flexio: Flexible Single- and Multi-Channel Speech Separation and Enhancement}, 
  year={2026},
  volume={},
  number={},
  pages={14417-14421},
  doi={10.1109/ICASSP55912.2026.11462393}}

@INPROCEEDINGS{sepdiff-chenbo,
  author={Chen, Bo and Wu, Chao and Zhao, Wenbin},
  booktitle={ICASSP 2023 - 2023 IEEE International Conference on Acoustics, Speech and Signal Processing (ICASSP)}, 
  title={SEPDIFF: Speech Separation Based on Denoising Diffusion Model}, 
  year={2023},
  volume={},
  number={},
  pages={1-5},
  doi={10.1109/ICASSP49357.2023.10095979}}

@INPROCEEDINGS{DiffSep-robin2023,
  author={Scheibler, Robin and Ji, Youna and Chung, Soo-Whan and Byun, Jaeuk and Choe, Soyeon and Choi, Min-Seok},
  booktitle={ICASSP 2023 - 2023 IEEE International Conference on Acoustics, Speech and Signal Processing (ICASSP)}, 
  title={Diffusion-Based Generative Speech Source Separation}, 
  year={2023},
  volume={},
  number={},
  pages={1-5},
  doi={10.1109/ICASSP49357.2023.10095310}}

@INPROCEEDINGS{SELM,
  author={Wang, Ziqian and Zhu, Xinfa and Zhang, Zihan and Lv, YuanJun and Jiang, Ning and Zhao, Guoqing and Xie, Lei},
  booktitle={ICASSP 2024 - 2024 IEEE International Conference on Acoustics, Speech and Signal Processing (ICASSP)}, 
  title={SELM: Speech Enhancement using Discrete Tokens and Language Models}, 
  year={2024},
  volume={},
  number={},
  pages={11561-11565},
  doi={10.1109/ICASSP48485.2024.10447464}}

@inproceedings{slmsurvey,
    title = "Recent Advances in Speech Language Models: A Survey",
    author = "Cui, Wenqian  and
      Yu, Dianzhi  and
      Jiao, Xiaoqi  and
      Meng, Ziqiao  and
      Zhang, Guangyan  and
      Wang, Qichao  and
      Guo, Steven Y.  and
      King, Irwin",
    editor = "Che, Wanxiang  and
      Nabende, Joyce  and
      Shutova, Ekaterina  and
      Pilehvar, Mohammad Taher",
    booktitle = "Proceedings of the 63rd Annual Meeting of the Association for Computational Linguistics (Volume 1: Long Papers)",
    month = jul,
    year = "2025",
    address = "Vienna, Austria",
    publisher = "Association for Computational Linguistics",
    url = "https://aclanthology.org/2025.acl-long.682/",
    doi = "10.18653/v1/2025.acl-long.682",
    pages = "13943--13970",
    ISBN = "979-8-89176-251-0"
}

@ARTICLE{valle,
  author={Chen, Sanyuan and Wang, Chengyi and Wu, Yu and Zhang, Ziqiang and Zhou, Long and Liu, Shujie and Chen, Zhuo and Liu, Yanqing and Wang, Huaming and Li, Jinyu and He, Lei and Zhao, Sheng and Wei, Furu},
  journal={IEEE Transactions on Audio, Speech and Language Processing}, 
  title={Neural Codec Language Models are Zero-Shot Text to Speech Synthesizers}, 
  year={2025},
  volume={33},
  number={},
  pages={705-718},
  doi={10.1109/TASLPRO.2025.3530270}}

@inproceedings{whisper,
  title={Robust speech recognition via large-scale weak supervision},
  author={Radford, Alec and Kim, Jong Wook and Xu, Tao and Brockman, Greg and McLeavey, Christine and Sutskever, Ilya},
  booktitle={International conference on machine learning},
  pages={28492--28518},
  year={2023},
  organization={PMLR}
}

@inproceedings{sot,
  title     = {Serialized Output Training for End-to-End Overlapped Speech Recognition},
  author    = {Naoyuki Kanda and Yashesh Gaur and Xiaofei Wang and Zhong Meng and Takuya Yoshioka},
  year      = {2020},
  booktitle = {Interspeech 2020},
  pages     = {2797--2801},
  doi       = {10.21437/Interspeech.2020-999},
  issn      = {2958-1796},
}

@INPROCEEDINGS{unisep,
  author={Wang, Yuanyuan and Chen, Hangting and Yang, Dongchao and Li, Weiqin and Luo, Dan and Li, Guangzhi and Yang, Shan and Wu, Zhiyong and Meng, Helen and Wu, Xixin},
  booktitle={2025 IEEE International Conference on Multimedia and Expo (ICME)}, 
  title={UniSep: Universal Target Audio Separation with Language Models at Scale}, 
  year={2025},
  volume={},
  number={},
  pages={1-6},
  doi={10.1109/ICME59968.2025.11210185}}

@ARTICLE{soundstream,
  author={Zeghidour, Neil and Luebs, Alejandro and Omran, Ahmed and Skoglund, Jan and Tagliasacchi, Marco},
  journal={IEEE/ACM Transactions on Audio, Speech, and Language Processing}, 
  title={SoundStream: An End-to-End Neural Audio Codec}, 
  year={2022},
  volume={30},
  number={},
  pages={495-507},
  doi={10.1109/TASLP.2021.3129994}}

@inproceedings{opensep,
    title = "{O}pen{S}ep: Leveraging Large Language Models with Textual Inversion for Open World Audio Separation",
    author = "Mahmud, Tanvir  and
      Marculescu, Diana",
    editor = "Al-Onaizan, Yaser  and
      Bansal, Mohit  and
      Chen, Yun-Nung",
    booktitle = "Proceedings of the 2024 Conference on Empirical Methods in Natural Language Processing",
    month = nov,
    year = "2024",
    address = "Miami, Florida, USA",
    publisher = "Association for Computational Linguistics",
    url = "https://aclanthology.org/2024.emnlp-main.735/",
    doi = "10.18653/v1/2024.emnlp-main.735",
    pages = "13244--13260",
}

@inproceedings{sepalm,
  title     = {SepALM: Audio Language Models Are Error Correctors for Robust Speech Separation},
  author    = {Mu, Zhaoxi and Yang, Xinyu and Wang, Gang},
  booktitle = {Proceedings of the Thirty-Fourth International Joint Conference on
               Artificial Intelligence, {IJCAI-25}},
  publisher = {International Joint Conferences on Artificial Intelligence Organization},
  editor    = {James Kwok},
  pages     = {8204--8212},
  year      = {2025},
  month     = {8},
  note      = {Main Track},
  doi       = {10.24963/ijcai.2025/912},
  url       = {https://doi.org/10.24963/ijcai.2025/912},
}

@inproceedings{tokensplit,
  title     = {TokenSplit: Using Discrete Speech Representations for Direct, Refined, and Transcript-Conditioned Speech Separation and Recognition},
  author    = {Hakan Erdogan and Scott Wisdom and Xuankai Chang and Zalán Borsos and Marco Tagliasacchi and Neil Zeghidour and John R. Hershey},
  year      = {2023},
  booktitle = {Interspeech 2023},
  pages     = {3462--3466},
  doi       = {10.21437/Interspeech.2023-2069},
  issn      = {2958-1796},
}

@inproceedings{sotasr,
  title     = {Adapting Multi-Lingual ASR Models for Handling Multiple Talkers},
  author    = {Chenda Li and Yao Qian and Zhuo Chen and Naoyuki Kanda and Dongmei Wang and Takuya Yoshioka and Yanmin Qian and Michael Zeng},
  year      = {2023},
  booktitle = {Interspeech 2023},
  pages     = {1314--1318},
  doi       = {10.21437/Interspeech.2023-1276},
  issn      = {2958-1796},
}

@INPROCEEDINGS{sotasr2,
  author={Kang, Jiawen and Meng, Lingwei and Cui, Mingyu and Guo, Haohan and Wu, Xixin and Liu, Xunying and Meng, Helen},
  booktitle={ICASSP 2024 - 2024 IEEE International Conference on Acoustics, Speech and Signal Processing (ICASSP)}, 
  title={Cross-Speaker Encoding Network for Multi-Talker Speech Recognition}, 
  year={2024},
  volume={},
  number={},
  pages={11986-11990},
  doi={10.1109/ICASSP48485.2024.10446249}}

@article{encodec,
title={High Fidelity Neural Audio Compression},
author={Alexandre D{\'e}fossez and Jade Copet and Gabriel Synnaeve and Yossi Adi},
journal={Transactions on Machine Learning Research},
issn={2835-8856},
year={2023},
note={Featured Certification, Reproducibility Certification}
}

@inproceedings{wavtokenizer,
 author = {Ji, Shengpeng and Jiang, Ziyue and Wang, Wen and Chen, Yifu and Fang, Minghui and Zuo, Jialong and Yang, Qian and Cheng, Xize and wang, zehan and Li, Ruiqi and Zhang, Ziang and Yang, Xiaoda and Huang, Rongjie and JIANG, YIDI and Chen, Qian and Zheng, Siqi and Zhao, Zhou},
 booktitle = {International Conference on Representation Learning},
 pages = {93809--93826},
 title = {WavTokenizer: an Efficient Acoustic Discrete Codec Tokenizer for Audio Language Modeling},
 volume = {2025},
 year = {2025}
}

@inproceedings{rvqgan,
 author = {Kumar, Rithesh and Seetharaman, Prem and Luebs, Alejandro and Kumar, Ishaan and Kumar, Kundan},
 booktitle = {Advances in Neural Information Processing Systems},
 editor = {A. Oh and T. Naumann and A. Globerson and K. Saenko and M. Hardt and S. Levine},
 pages = {27980--27993},
 publisher = {Curran Associates, Inc.},
 title = {High-Fidelity Audio Compression with Improved RVQGAN},
 url = {https://proceedings.neurips.cc/paper_files/paper/2023/file/58d0e78cf042af5876e12661087bea12-Paper-Conference.pdf},
 volume = {36},
 year = {2023}
}

@INPROCEEDINGS{lisa,
  author={Huang, Jiankai and Zhang, Junteng and Lu, Ming and Cao, Xun and Ma, Zhan},
  booktitle={ICASSP 2026 - 2026 IEEE International Conference on Acoustics, Speech and Signal Processing (ICASSP)}, 
  title={Lisa: Lightweight Yet Superb Neural Speech Coding}, 
  year={2026},
  volume={},
  number={},
  pages={14457-14461},
  doi={10.1109/ICASSP55912.2026.11464260}}

@article{wavlm,
  title={Wavlm: Large-scale self-supervised pre-training for full stack speech processing},
  author={Chen, Sanyuan and Wang, Chengyi and Chen, Zhengyang and Wu, Yu and Liu, Shujie and Chen, Zhuo and Li, Jinyu and Kanda, Naoyuki and Yoshioka, Takuya and Xiao, Xiong and others},
  journal={IEEE Journal of Selected Topics in Signal Processing},
  volume={16},
  number={6},
  pages={1505--1518},
  year={2022},
  publisher={IEEE}
}

@INPROCEEDINGS{sepeda,
  author={Lee, Younglo and Choi, Shukjae and Kim, Byeong-Yeol and Wang, Zhong-Qiu and Watanabe, Shinji},
  booktitle={ICASSP 2024 - 2024 IEEE International Conference on Acoustics, Speech and Signal Processing (ICASSP)}, 
  title={Boosting Unknown-Number Speaker Separation with Transformer Decoder-Based Attractor}, 
  year={2024},
  volume={},
  number={},
  pages={446-450},
  doi={10.1109/ICASSP48485.2024.10446032}}

@ARTICLE{SD-SepEDA,
  author={Chetupalli, Srikanth Raj and Habets, Emanuël A. P.},
  journal={IEEE/ACM Transactions on Audio, Speech, and Language Processing}, 
  title={Speaker Counting and Separation From Single-Channel Noisy Mixtures}, 
  year={2023},
  volume={31},
  number={},
  pages={1681-1692},
  doi={10.1109/TASLP.2023.3268572}}

@article{eendeda,
  title={Encoder-decoder based attractors for end-to-end neural diarization},
  author={Horiguchi, Shota and Fujita, Yusuke and Watanabe, Shinji and Xue, Yawen and Garcia, Paola},
  journal={IEEE/ACM Transactions on Audio, Speech, and Language Processing},
  volume={30},
  pages={1493--1507},
  year={2022},
  publisher={IEEE}
}

@INPROCEEDINGS{MUSE,
  author={Saijo, Kohei and Zhang, Wangyou and Wang, Zhong-Qiu and Watanabe, Shinji and Kobayashi, Tetsunori and Ogawa, Tetsuji},
  booktitle={2023 IEEE Automatic Speech Recognition and Understanding Workshop (ASRU)}, 
  title={A Single Speech Enhancement Model Unifying Dereverberation, Denoising, Speaker Counting, Separation, And Extraction}, 
  year={2023},
  volume={},
  number={},
  pages={1-6},
  doi={10.1109/ASRU57964.2023.10389724}}

@INPROCEEDINGS{promptsep,
  author={Wen, Yutong and Chen, Ke and Seetharaman, Prem and Nieto, Oriol and Su, Jiaqi and Kumar, Rithesh and Kim, Minje and Smaragdis, Paris and Jin, Zeyu and Salamon, Justin},
  booktitle={ICASSP 2026 - 2026 IEEE International Conference on Acoustics, Speech and Signal Processing (ICASSP)}, 
  title={PromptSep: Generative Audio Separation Via Multimodal Prompting}, 
  year={2026},
  volume={},
  number={},
  pages={14422-14426},
  doi={10.1109/ICASSP55912.2026.11460803}}

@misc{librimix,
      title={LibriMix: An Open-Source Dataset for Generalizable Speech Separation}, 
      author={Joris Cosentino and Manuel Pariente and Samuele Cornell and Antoine Deleforge and Emmanuel Vincent},
      year={2020},
      eprint={2005.11262},
      archivePrefix={arXiv},
      primaryClass={eess.AS},
}

@inproceedings{pit,
  title={Permutation invariant training of deep models for speaker-independent multi-talker speech separation},
  author={Yu, Dong and Kolb{\ae}k, Morten and Tan, Zheng-Hua and Jensen, Jesper},
  booktitle={2017 IEEE International Conference on Acoustics, Speech and Signal Processing (ICASSP)},
  pages={241--245},
  year={2017},
  organization={IEEE}
}

@ARTICLE{pit2,
  author={Kolbæk, Morten and Yu, Dong and Tan, Zheng-Hua and Jensen, Jesper},
  journal={IEEE/ACM Transactions on Audio, Speech, and Language Processing}, 
  title={Multitalker Speech Separation With Utterance-Level Permutation Invariant Training of Deep Recurrent Neural Networks}, 
  year={2017},
  volume={25},
  number={10},
  pages={1901-1913},
  doi={10.1109/TASLP.2017.2726762}}

@INPROCEEDINGS{lps,
  author={Pirklbauer, Jan and Sach, Marvin and Fluyt, Kristoff and Tirry, Wouter and Wardah, Wafaa and Moeller, Sebastian and Fingscheidt, Tim},
  booktitle={Speech Communication; 15th ITG Conference}, 
  title={Evaluation Metrics for Generative Speech Enhancement Methods: Issues and Perspectives}, 
  year={2023},
  volume={},
  number={},
  pages={265-269},
  doi={10.30420/456164052}}

@inproceedings{speechbertscore,
  title={SpeechBERTScore: Reference-Aware Automatic Evaluation of Speech Generation Leveraging NLP Evaluation Metrics},
  author={Saeki, Takaaki and Maiti, Soumi and Takamichi, Shinnosuke and Watanabe, Shinji and Saruwatari, Hiroshi},
  booktitle={Proc. Interspeech 2024},
  pages={4943--4947},
  year={2024}
}

@article{fireredasr-xu2025,
  title={Fireredasr: Open-source industrial-grade mandarin speech recognition models from encoder-decoder to llm integration},
  author={Xu, Kai-Tuo and Xie, Feng-Long and Tang, Xu and Hu, Yao},
  journal={arXiv preprint arXiv:2501.14350},
  year={2025}
}

@article{seedasr-bai2024,
  title={Seed-asr: Understanding diverse speech and contexts with llm-based speech recognition},
  author={Bai, Ye and Chen, Jingping and Chen, Jitong and Chen, Wei and Chen, Zhuo and Ding, Chuang and Dong, Linhao and Dong, Qianqian and Du, Yujiao and Gao, Kepan and others},
  journal={arXiv preprint arXiv:2407.04675},
  year={2024}
}

@article{seedtts-anastassiou2024,
  title={Seed-TTS: A Family of High-Quality Versatile Speech Generation Models},
  author={Seed Team, ByteDance},
  journal={arXiv preprint arXiv:2406.02430},
  year={2024}
}

@INPROCEEDINGS{mossformer2,
  author={Zhao, Shengkui and Ma, Yukun and Ni, Chongjia and Zhang, Chong and Wang, Hao and Nguyen, Trung Hieu and Zhou, Kun and Yip, Jia Qi and Ng, Dianwen and Ma, Bin},
  booktitle={ICASSP 2024 - 2024 IEEE International Conference on Acoustics, Speech and Signal Processing (ICASSP)}, 
  title={MossFormer2: Combining Transformer and RNN-Free Recurrent Network for Enhanced Time-Domain Monaural Speech Separation}, 
  year={2024},
  volume={},
  number={},
  pages={10356-10360},
  doi={10.1109/ICASSP48485.2024.10445985}}

@INPROCEEDINGS{pesq,
  author={Rix, A.W. and Beerends, J.G. and Hollier, M.P. and Hekstra, A.P.},
  booktitle={2001 IEEE International Conference on Acoustics, Speech, and Signal Processing. Proceedings (Cat. No.01CH37221)}, 
  title={Perceptual evaluation of speech quality (PESQ)-a new method for speech quality assessment of telephone networks and codecs}, 
  year={2001},
  volume={2},
  number={},
  pages={749-752 vol.2},
  doi={10.1109/ICASSP.2001.941023}}

@INPROCEEDINGS{dnsmos,
  author={Reddy, Chandan K A and Gopal, Vishak and Cutler, Ross},
  booktitle={ICASSP 2021 - 2021 IEEE International Conference on Acoustics, Speech and Signal Processing (ICASSP)}, 
  title={Dnsmos: A Non-Intrusive Perceptual Objective Speech Quality Metric to Evaluate Noise Suppressors}, 
  year={2021},
  volume={},
  number={},
  pages={6493-6497},
  doi={10.1109/ICASSP39728.2021.9414878}}

@inproceedings{utmos,
  title     = {{UTMOS: UTokyo-SaruLab System for VoiceMOS Challenge 2022}},
  author    = {Takaaki Saeki and Detai Xin and Wataru Nakata and Tomoki Koriyama and Shinnosuke Takamichi and Hiroshi Saruwatari},
  year      = {2022},
  booktitle = {{Interspeech 2022}},
  pages     = {4521--4525},
  doi       = {10.21437/Interspeech.2022-439},
  issn      = {2958-1796},
}

@INPROCEEDINGS{tflocoformer,
  author={Saijo, Kohei and Wichern, Gordon and Germain, François G. and Pan, Zexu and Roux, Jonathan Le},
  booktitle={2024 18th International Workshop on Acoustic Signal Enhancement (IWAENC)}, 
  title={TF-Locoformer: Transformer with Local Modeling by Convolution for Speech Separation and Enhancement}, 
  year={2024},
  volume={},
  number={},
  pages={205-209},
  doi={10.1109/IWAENC61483.2024.10694313}}

@INPROCEEDINGS{aishell1,
  author={Bu, Hui and Du, Jiayu and Na, Xingyu and Wu, Bengu and Zheng, Hao},
  booktitle={2017 20th Conference of the Oriental Chapter of the International Coordinating Committee on Speech Databases and Speech I/O Systems and Assessment (O-COCOSDA)}, 
  title={AISHELL-1: An open-source Mandarin speech corpus and a speech recognition baseline}, 
  year={2017},
  volume={},
  number={},
  pages={1-5},
  doi={10.1109/ICSDA.2017.8384449}}

@inproceedings{wham,
  title     = {{WHAM!: Extending Speech Separation to Noisy Environments}},
  author    = {Gordon Wichern and Joe Antognini and Michael Flynn and Licheng Richard Zhu and Emmett McQuinn and Dwight Crow and Ethan Manilow and Jonathan Le Roux},
  year      = {2019},
  booktitle = {{Interspeech 2019}},
  pages     = {1368--1372},
  doi       = {10.21437/Interspeech.2019-2821},
  issn      = {2958-1796},
}

\section{Author Biography}
\begin{IEEEbiography}[{\includegraphics[width=1in,height=1.25in,clip]{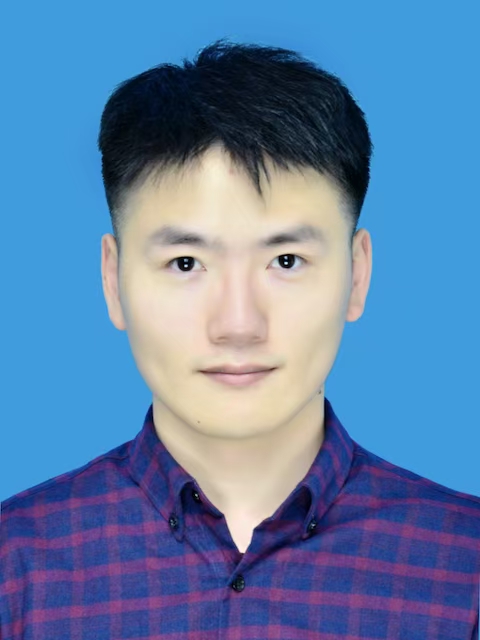}}]{Chenda Li}
(Member, IEEE) received the B.Eng. degree in electronic information engineering from Huazhong University of Science and Technology, Wuhan, China, in 2018, and the M.S. and Ph.D. degrees in computer science and technology from Shanghai Jiao Tong University, Shanghai, China, in 2020 and 2024, respectively. 
He was a Visiting Scholar at Carnegie Mellon University from 2023 to 2024.
He is currently an Assistant Professor with the AudioCC Lab, School of Computer Science, Shanghai Jiao Tong University. He has authored or coauthored more than 50 papers in peer-reviewed journals and conferences, including T-ASLP, ICASSP, INTERSPEECH, SLT, and ASRU. He is a core developer and maintainer of ESPnet-SE, an open-source speech enhancement and separation toolkit, and a co-organizer of the URGENT speech enhancement challenge series. He was selected for the 2025 Microsoft Research Asia StarTrack Program. His research interests include speech enhancement and separation, sound separation, multimodal speech and audio processing, speech language models, and generative audio models.
\end{IEEEbiography}

\begin{IEEEbiography}[{\includegraphics[width=1in,height=1.25in, clip]{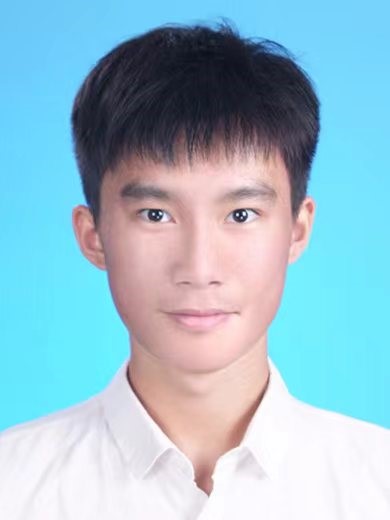}}]{Tianhua Li} (Student Member, IEEE) received the B.Eng. degree from the School of Electronic Information and Electrical Engineering, Shanghai Jiao Tong University, Shanghai, China, in 2025. He is currently working toward the M.Sc. degree with the AudioCC Lab, School of Computer Science, Shanghai Jiao Tong University, Shanghai, China, under the supervision of Yanmin Qian. His current research interests include Sound Separation, Speech Language Model, and Audio Editing.

\end{IEEEbiography}

\begin{IEEEbiography}[{\includegraphics[width=1in,height=1.25in, clip]{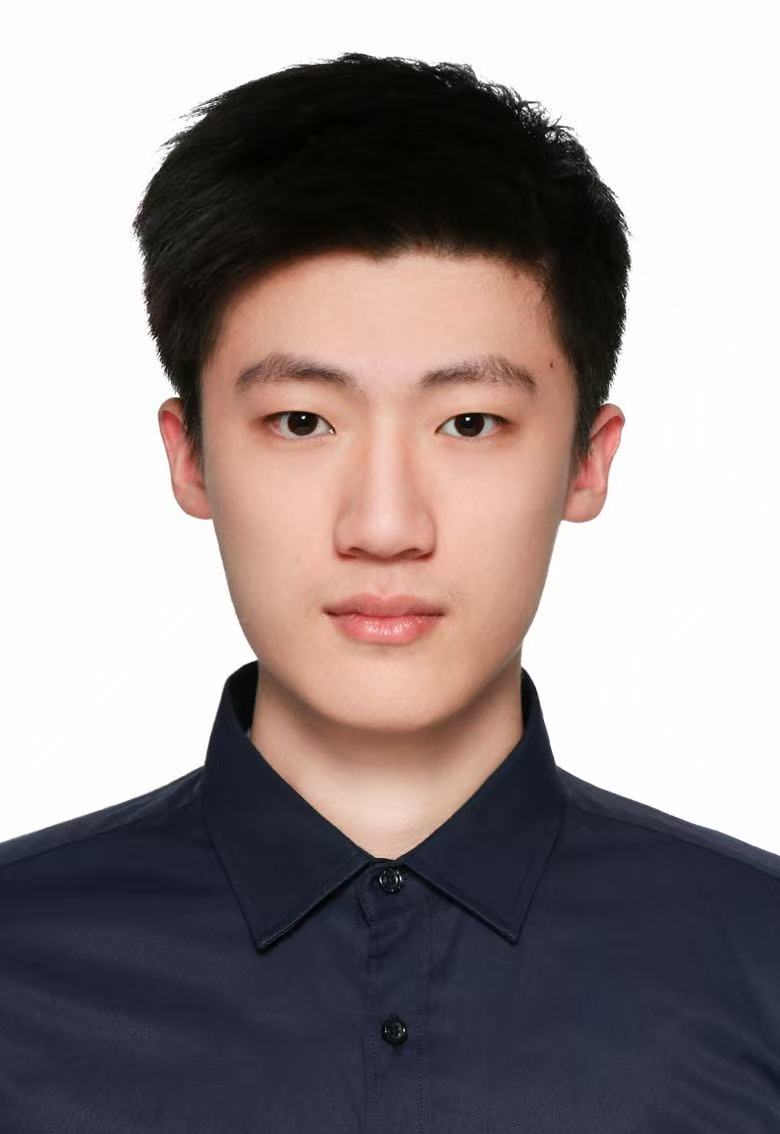}}]{Yiming Chen}
is currently pursuing the Ph.D. degree with the School of Computer Science, Shanghai Jiao Tong University, Shanghai, China, where he received the B.E. degree in computer science in 2026. He conducts his research at the Auditory Cognition and Computational Acoustics (AudioCC) Lab, led by Prof. Yanmin Qian. His research interests include machine learning and its applications, particularly speech language models (SLMs), with current work focusing on multi-speaker speech recognition and speech separation.
\end{IEEEbiography}

\begin{IEEEbiography}[{\includegraphics[width=1in,height=1.25in,clip]{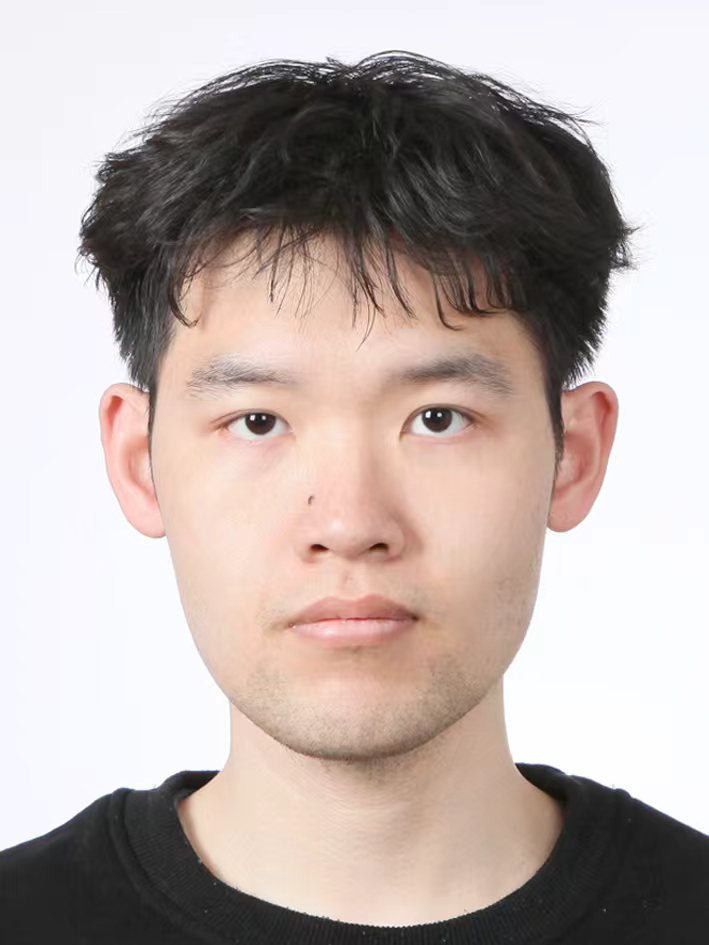}}]{Wei Wang}
received the the Ph.D. degree in computer science and technology from Shanghai Jiao Tong University, Shanghai, China, in 2026, under the supervision of Yanmin Qian. His research interests include automatic speech recognition, spoken-language understanding, speech synthesis, speech enhancement, and speech-quality assessment.
\end{IEEEbiography}

\begin{IEEEbiography}
[{\includegraphics[width=1in,height=1.25in, clip]{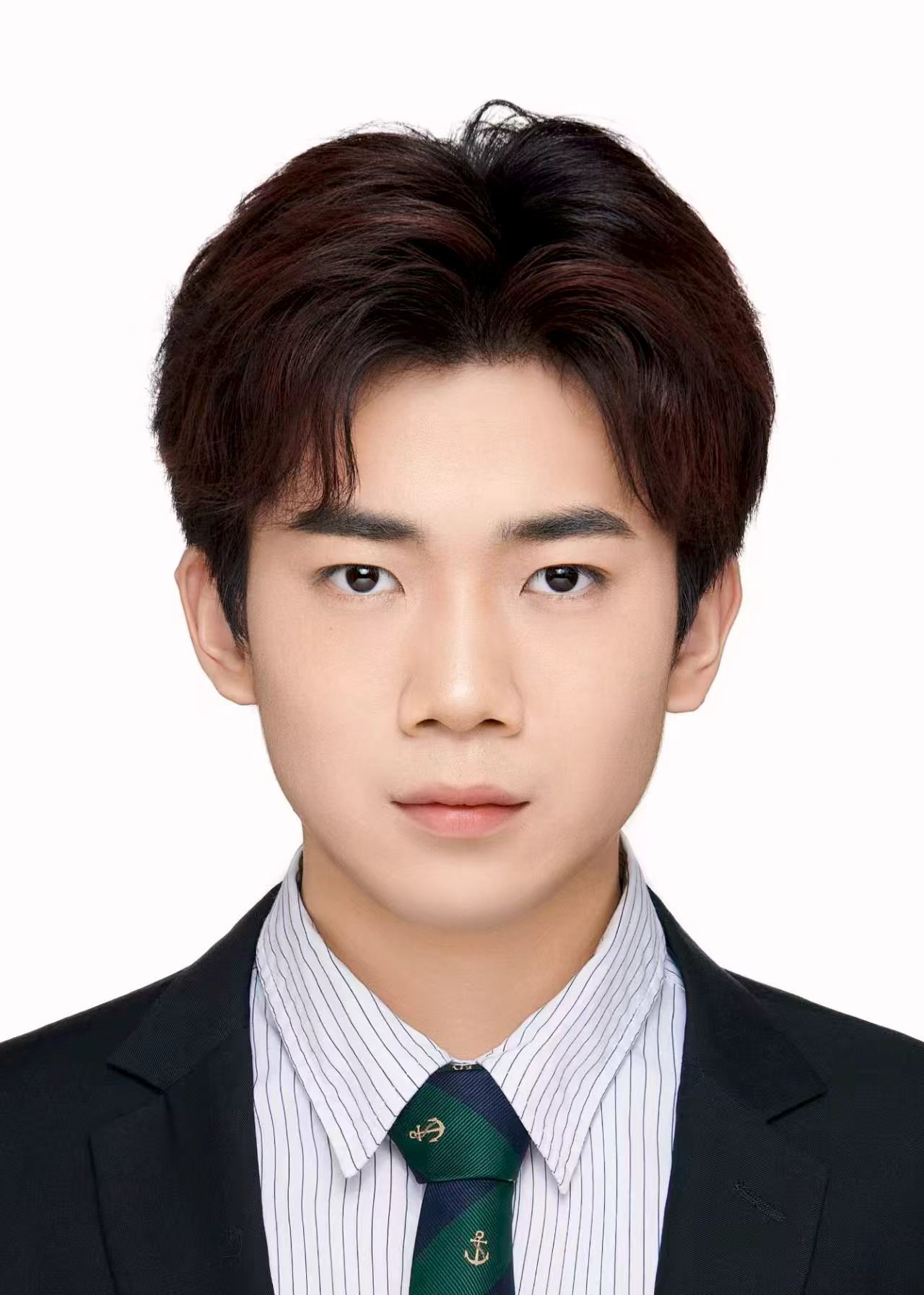}}]{Jiahe Wang} (Student Member, IEEE) received the B.Eng. degree from the Faculty of Electronic and Information Engineering, Xi'an Jiaotong University, Xi'an, China, in 2024, and is currently working toward the M.Sc. degree with the AudioCC Lab, School of Computer Science, Shanghai Jiao Tong University, Shanghai, China, under the supervision of Professor Yanmin Qian. His current research interests include speech enhancement (SE) and speech quality assessment (SQA).
\end{IEEEbiography}

\begin{IEEEbiography}[{\includegraphics[width=1in,height=1.25in, clip]{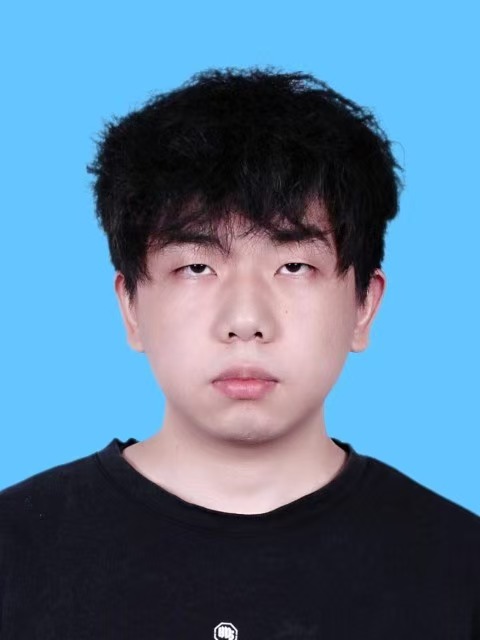}}]{Xin Zhou} received the B.Eng. degree in Computer Science and Technology from the University of Electronic Science and Technology of China, Chengdu, China, in 2023, and the Master's degree in Electronic and Information Engineering from Shanghai Jiao Tong University, Shanghai, China, in 2026, under the supervision of Yanmin Qian. He is currently an engineer at the Shanghai Artificial Intelligence Laboratory, Shanghai, China. His research interests include speech separation.
\end{IEEEbiography}

\begin{IEEEbiography}[{\includegraphics[width=1in,height=1.25in, clip]{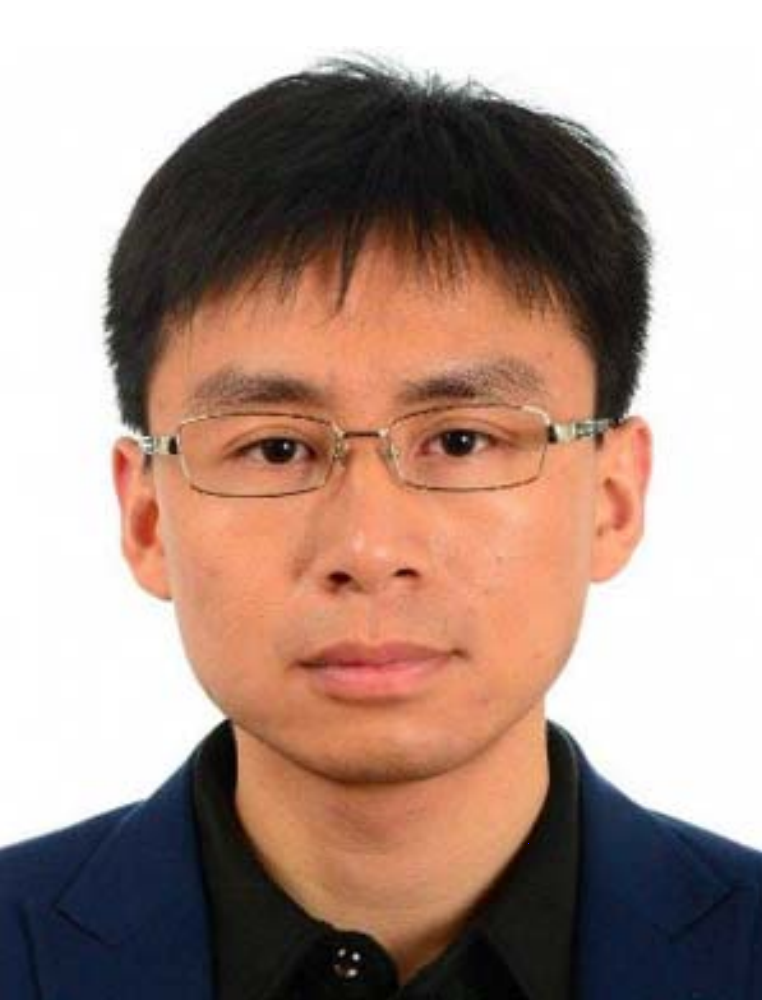}}]{Yanmin Qian} (Senior Member, IEEE) received the B.S. degree from the Department of Electronic and Information Engineering, Huazhong University of Science and Technology, Wuhan, China, in 2007, and the Ph.D. degree from the Department of Electronic Engineering, Tsinghua University, Beijing, China, in 2012. Since 2013, he has been with the School of Computer Science, Shanghai Jiao Tong University, Shanghai, China, where he is currently a Full Professor. From 2015 to 2016, he was an Associate Research with the Speech Group, Cambridge University Engineering Department, Cambridge, U.K. He has authored or coauthored more than 300 papers in peer-reviewed journals and conferences on speech and language processing, including T-ASLP, Speech Communication, ICASSP, INTERSPEECH and ASRU. He has applied for more than 120 Chinese and American patents and won 6 championships of international challenges. His current research interests include speech and audio signal processing, automatic speech recognition and translation, speaker and language recognition, speech separation and enhancement, music generation and understanding, speech emotion perception, multimodal information processing, natural language understanding, deep learning and multi-media signal processing. He was the recipient of several top academic awards in China, including Chang Jiang Scholars Program of the Ministry of Education, Excellent Youth Fund of the National Natural Science Foundation of China, and the First Prize of Wu Wenjun Artificial Intelligence Science and Technology Award (First Completion). He was also the recipient of several awards from international research committee, including the Best Paper Awards in IEEE Transactions on Audio, Speech and Language Processing (2025), Speech Communication (2019) and Best Paper Awards from IEEE ISCSLP 2024, IEEE DCASE 2024, IEEE ASRU 2019, IEEE ISCSLP 2016. He is also the Member of IEEE Signal Processing Society Speech and Language Technical Committee.
\end{IEEEbiography}

\end{document}